\def\E{\end{document}}
\documentclass[10pt]{article}
\usepackage{amssymb,amsmath}

\begin{document}
\title{
Critical line of exponents, scattering theories for a weighted gradient system of semilinear wave equations
}
 \author{Xiaowei An$^{1,2}${\thanks{E-mail: anxiaowei2024@163.com (X. W. An).
}}\\
\small 1 School of Intelligence Policing, China People's Police University,\\
\small Langfang, He Bei, 065000, P. R. China\\
\small 2 Hebei Key Laboratory of Information Support Technology for Smart Policing,\\
\small China People's Police University, Langfang 065000, P. R. China\\
Xianfa Song$^{3,4}${\thanks{E-mail:\ \tt songxianfa@tju.edu.cn (or songxianfa2004@163.com)
  }}\\
\small 3 Department of Mathematics, School of Mathematics, Tianjin University,\\
\small Tianjin, 300072, P. R. China\\
\small 4 Xinjiang Production and Construction Corps\\
\small Key Laboratory of Green and Intelligent Development and Efficient Utilization of Strategic Mineral Resources,\\
\small Xinjiang University of Technology,  Hotan  Xinjiang 84800, P.R. China\\
}

\maketitle
\date{}

\newtheorem{theorem}{Theorem}[section]
\newtheorem{definition}{Definition}[section]
\newtheorem{lemma}{Lemma}[section]
\newtheorem{proposition}{Proposition}[section]
\newtheorem{corollary}{Corollary}[section]
\newtheorem{remark}{Remark}[section]
\renewcommand{\theequation}{\thesection.\arabic{equation}}
\catcode`@=11 \@addtoreset{equation}{section} \catcode`@=12

\begin{abstract}

\quad In this paper, we consider the following Cauchy problem of a weighted gradient system of semilinear wave equations
\begin{equation*}
\left\{
\begin{array}{lll}
u_{tt}-\Delta u=\lambda |u|^{\alpha}|v|^{\beta+2}u,\quad v_{tt}-\Delta v=\mu |u|^{\alpha+2}|v|^{\beta}v,\quad x\in \mathbb{R}^d,\ t\in \mathbb{R},\\
u(x,0)=u_{10}(x),\ u_t(x,0)=u_{20}(x),\quad v(x,0)=v_{10}(x),\ v_t(x,0)=v_{20}(x),\quad x\in \mathbb{R}^d.
\end{array}\right.
\end{equation*}
Here $d\geq 3$, $\lambda, \mu\in \mathbb{R}$, $\alpha, \beta\geq 0$, $(u_{10},u_{20})$ and $(v_{10},v_{20})$ belong to $H^1(\mathbb{R}^d)\oplus L^2(\mathbb{R}^d)$ or $\dot{H}^1(\mathbb{R}^d)\oplus L^2(\mathbb{R}^d)$
or $\dot{H}^{\gamma}(\mathbb{R}^d)\oplus H^{\gamma-1}(\mathbb{R}^d)$ for some $\gamma>1$. Under certain assumptions, we establish the local wellposedness of the $H^1\oplus H^1$-solution, $\dot{H}^1\oplus \dot{H}^1$-solution and $\dot{H}^{\gamma}\oplus \dot{H}^{\gamma}$-solution of the system with different types of initial data.



\end{abstract}

{\bf Contents.}

{\bf 1 Introduction.}

{\bf 2 Preliminaries.}

{\bf 3 Global wellpossedness results on  (\ref{1}) when $d=3$ and $\alpha+\beta\leq 2$.}

{\bf 4 Regular results on (\ref{1}) in some special cases.}

{\bf 5 $\dot{H}^1\oplus \dot{H}^1$ scattering theory for (\ref{1}) in some special cases.}

{\bf 6 $\dot{H}^{s_c}\oplus \dot{H}^{s_c}$ scattering theory for (\ref{1}) in the supercritical case.}

{\bf References.}

\section{Introduction}
\qquad In this paper, we consider the following Cauchy problem
\begin{equation}
\label{1}\left\{
\begin{array}{lll}
u_{tt}-\Delta u=\lambda |u|^{\alpha}|v|^{\beta+2}u,\quad v_{tt}-\Delta v=\mu |u|^{\alpha+2}|v|^{\beta}v,\quad x\in \mathbb{R}^d,\ t\in \mathbb{R},\\
u(x,0)=u_{10}(x),\ u_t(x,0)=u_{20}(x),\quad v(x,0)=v_{10}(x),\ v_t(x,0)=v_{20}(x),\quad x\in \mathbb{R}^d.
\end{array}\right.
\end{equation}
Here $d\geq 2$, $\lambda, \mu\in \mathbb{R}$, $\alpha, \beta\geq 0$, $(u_{10},u_{20})$ and $(v_{10},v_{20})$ belong to $H^1(\mathbb{R}^d)\oplus L^2(\mathbb{R}^d)$ or $\dot{H}^1(\mathbb{R}^d)\oplus L^2(\mathbb{R}^d)$
or $\dot{H}^{\gamma}(\mathbb{R}^d)\oplus H^{\gamma-1}(\mathbb{R}^d)$ for some $\gamma>1$. Besides the local wellposedness of the solution, we are concerned with the global existence, regularity, asymptotic behavior and scattering theory for (\ref{1}) in this paper.

A large amount of work had been devoted to the scattering theory
for the following Cauchy problem
\begin{equation}
\label{1'}\left\{
\begin{array}{lll}
u_{tt}-\Delta u=\pm |u|^pu,\quad x\in \mathbb{R}^d,\ t\in \mathbb{R},\\
u(x,0)=u_{10}(x),\ u_t(x,0)=u_{20}(x),\quad x\in \mathbb{R}^d.
\end{array}\right.
\end{equation}
It is well known that $p=\frac{4}{n-2}$ is said to be a critical
exponent of the $H^1$ solution of (\ref{1'}). The global existence and scattering theory in different energy space in subcritical case can be seen in \cite{Dodson, Dodson2015, Ginibre1989, Ginibre19892, Pecher1988, Pecher1984, Shen, Strauss, Struwe}), while  parallel results on (\ref{1'}) in the critical case can be seen in \cite{Bahouri, Donninger, Duyckaerts, Kenig, Krieger} and
those  on (\ref{1'}) in the super-critical case can be seen in \cite{Bulut, Gao2020, Kenig2011, Killip}. For the existence and scattering of solutions with small data at lower regularity, we can refer to \cite{ Kapitanskii, Lindblad, Pecher1996, Wang}. For almost--sure scattering for the radial energy--critical nonlinear wave equation, we can refer to \cite{Bringmann, Dodson}. Some authors dealt with  the regularity and asymptotic behavior for the wave equation with a critical
nonlinearity(see \cite{Grillakis, Lindblad, Shatah1993, Shatah1994}).

There are also many lectures on scattering theory for Klein-Gordon equation, see \cite{Brenner, Ginibre1985,Ginibre1989, Nakanishi19992, Pecher1984, Pecher1985, Struwe1988}. Meanwhile, the asymptotic behavior for Klein-Gordon equation in Schwarzschild metric can be seen in \cite{Bachelot,Dimock}.

Some results on the wave equations with other type of nonlinearities were established. Global solutions to the wave
equations with nonlinearity of exponential growth in the critical Sobolev space was established in \cite{Nakamura}.  Unique global existence and asymptotic behavior for solutions to wave equations with non-coercive nonlinearity was established in \cite{Nakanishi1999}. Wave equations with time-dependent dissipation was discussed in \cite{Wirth}. Random data Cauchy theory for nonlinear wave equations was studied in \cite{Luhrmann}.

To analyze the behavior for the solutions to wave equations, different type of estimates were established(see \cite{Beals, Cote, Ginibre1995, Marshall, Pecher1976, Peral, Strichartz1, Strichartz2, Tataru}). We can refer to these books \cite{Deift, Jeffrey, Lax, Stein, Tao} and see more information on wave equations.

About the results on the global solution and scattering theory for a system of wave equations, we can refer to \cite{Del, Katayama, Kubo2003, Kubo2002} and the references therein. Differing to the equations have the forms of $u_{tt}-\Delta u=H_v(u,v)$ and $v_{tt}-\Delta v=H_u(u,v)$ in these references above,
 the equations in (\ref{1}) have their special structure as follows: There exist two constants $A=\mu(\alpha+2)$, $B=\lambda(\beta+2)$ and a function $F(u,v)=\lambda\mu|u|^{\alpha+2}|v|^{\beta+2}$  such that
\begin{align}
\frac{\partial F(u,v)}{\partial u}=A\lambda  |u|^{\alpha}|v|^{\beta+2}u,\quad \frac{\partial F(u,v)}{\partial v}=B \mu |u|^{\alpha+2}|v|^{\beta}v,\label{221051}
\end{align}
which will be called as a weighted(essential) gradient system of wave equations below. We will show an interesting phenomenon on (\ref{1}) below. There exists a critical line of exponents $\alpha+\beta=2$ when $d=3$ in the following sense: The system always has a unique bounded $H^1\oplus H^1$-solution for any initial data $(u_{10},u_{20}), (v_{10},v_{20})\in H^1(\mathbb{R}^3)\oplus L^2(\mathbb{R}^3)$ if $\alpha+\beta\leq 2$, while it fails to be well-posed in $H^1(\mathbb{R}^3)\oplus H^1(\mathbb{R}^3)$ for some $(\alpha,\beta)$ satisfying $\alpha+\beta>2$. While when $d=4$, we call $(\alpha,\beta)=(0,0)$ is the critical point of exponents in the following sense: The system always has a unique bounded $H^1\oplus H^1$-solution for any initial data $(u_{10},u_{20}), (v_{10},v_{20})\in H^1(\mathbb{R}^4)\oplus L^2(\mathbb{R}^4)$ if $(\alpha,\beta)=(0,0)$, while it fails to be well-posed in $H^1(\mathbb{R}^4)\oplus H^1(\mathbb{R}^4)$ if $\alpha+\beta>0$. We can define the weighted energy of (\ref{1}) as follows.
\begin{align}
E_w(u,v)=A(\|\nabla u\|^2_2+\|u_t\|^2_2)+B(\|\nabla v\|^2_2+\|v_t\|^2_2)+\lambda\mu\int_{\mathbb{R}^d}|u|^{\alpha+2}|v|^{\beta+2}dx.\label{223s1}
\end{align}
Based on the conservation of weighted energy, we first get the global existence result, which is one of main purposes in this paper. Our second goal is to obtain the regularity results on (\ref{1}). Last, we will establish scattering theory for (\ref{1}) in supercritical case.

The rest of this paper is organized as follows. In Section 2, we give some notations and useful lemmas. In Section 3, we will consider the global wellposedness of (\ref{1}) when $d=2$ and $d=3$. In Section 4, we will get the regularity results on (\ref{1}). In Section 5, we will eatablish $\dot{H}^1\oplus \dot{H}^1$ scattering theory for (\ref{1}) in some special cases. In the last section, we will establish scattering theory for (\ref{1}) in supercritical case.

\section{Preliminaries}

\qquad In this section, we give some notations and lemmas.

{\bf Definition 2.1.} {\it If there exist two constants $A$, $B$ and a function $F(u,v)$  such that
\begin{align}
\frac{\partial F(u,v)}{\partial u}=Af_1(u,v),\quad \frac{\partial F(u,v)}{\partial v}=Bf_2(u,v),\label{221051}
\end{align}
then we call the system
\begin{align}
u_{tt}-\Delta u=f_1(u,v),\quad v_{tt}-\Delta v=f_2(u,v)
\end{align}
 as a {\bf weighted(essential) gradient system of wave equations}.}

Let $F(u,v)=\lambda \mu |u|^{\alpha+2}|v|^{\beta+2}$, $A=\mu(\alpha+2)$ and $B=\lambda(\beta+2)$. Then
\begin{align}
\frac{\partial F(u,v)}{\partial u}=Af_1(u,v)=A\lambda  |u|^{\alpha}|v|^{\beta+2}u,\quad \frac{\partial F(u,v)}{\partial u}=Bf_2(u,v)=B \mu |u|^{\alpha+2}|v|^{\beta}v,
\end{align}
which means that the system $u_{tt}-\Delta u=\lambda  |u|^{\alpha}|v|^{\beta+2}u$, $v_{tt}-\Delta v= \mu |u|^{\alpha+2}|v|^{\beta}v$  is a  weighted(essential) gradient system of wave equations.

We denote by $\mathcal{F}$ Fourier transform, and we write  $\hat{v}=\mathcal{F}v$ for any $v\in \mathcal{J}'(\mathbb{R}^d)$. Define the function spaces
\begin{align}
H^s_p&=H^s_p(\mathbb{R}^d)=\{f\in \mathcal{J}'(\mathbb{R}^d):\mathcal{F}^{-1}[(1+|\xi|^2)^{\frac{s}{2}}\hat{f}]\in L^p(\mathbb{R}^d) \}
\end{align}
equipped with the norm $\|f\|_{H^s_p}=\|\mathcal{F}^{-1}[(1+|\xi|^2)^{\frac{s}{2}}\hat{f}]\|_{L^p}$ and
\begin{align}
\dot{H}^s_p&=\dot{H}^s_p(\mathbb{R}^d)=\{f\in \mathcal{J}'(\mathbb{R}^d):\mathcal{F}^{-1}[|\xi|^s\hat{f}]\in L^p(\mathbb{R}^d) \},
\end{align}
equipped with the norm $\|f\|_{\dot{H}^s_p}=\|\mathcal{F}^{-1}[|\xi|^s\hat{f}]\|_{L^p}$

Define the operators $\omega=(-\Delta)^{\frac{1}{2}}$, $K(t)=\omega^{-1}\sin (\omega t)$ and $\dot{K}=\cos (\omega t)$.
Denote
\begin{equation}
\label{7281}\left\{
\begin{array}{lll}
G_1(t_1,t_2;u,v)(t)=\int_{t_1}^{t_2}K(t-\tau)(\lambda |u(\tau)|^{\alpha}|v(\tau)|^{\beta+2}u(\tau))d\tau,\\
 G_2(t_1,t_2;u,v)(t)=\int_{t_1}^{t_2}K(t-\tau)(\mu |u(\tau)|^{\alpha+2}|v(\tau)|^{\beta}v(\tau))d\tau,\\
F_1(t_0;u,v)(t)=G_1(t_0,t;u,v)(t),\quad F_2(t_0;u,v)(t)=G_2(t_0,t;u,v)(t),\\
F_1(t_2;u,v)-F_1(t_1;u,v)=G_1(t_1,t_2;u,v),\\
F_2(t_2;u,v)-F_2(t_1;u,v)=G_2(t_1,t_2;u,v).
\end{array}\right.
\end{equation}

We will construct a local(in $t$) solution to the integral equation
\begin{equation}
\label{771}\left\{
\begin{array}{lll}
u(t)=\dot{K}(t)u_{10}+K(t)u_{20}+\int_0^t(-\Delta)^{-\frac{1}{2}}\sin[(-\Delta)^{\frac{1}{2}}(t-\tau)](\lambda |u|^{\alpha}|v|^{\beta+2}u)d\tau,\\
\qquad :=\dot{K}(t)u_{10}+K(t)u_{20}+F_1(t;u,v)=A_1(u_{10},u_{20}),\\
v(t)=\dot{K}(t)v_{10}+K(t)v_{20}+\int_0^t(-\Delta)^{-\frac{1}{2}}\sin[(-\Delta)^{\frac{1}{2}}(t-\tau)](\mu |u|^{\alpha+2}|v|^{\beta}v)d\tau,\\
\qquad :=\dot{K}(t)v_{10}+K(t)v_{20}+F_2(t;u,v)=A_2(v_{10},v_{20}).
\end{array}\right.
\end{equation}

Let
\begin{align}
\frac{2\eta(r)}{(d+1)}=\frac{\gamma(r)}{(d-1)}=\frac{\delta(r)}{d}=\frac{1}{2}-\frac{1}{r}.\label{4161}
\end{align}

We will recall some properties for the operators $\omega$ and $K(t)$ below(see \cite{Marshall, Peral, Strichartz1}).

{\bf Lemma 2.1.} {\it
(i) For any $\psi\in C^{\infty}_0(\mathbb{R}^d)$ and $t>0$, $1<p\leq 2 \leq q<\infty$,
\begin{align}
\|K(t)\psi\|_{L^{q}(\mathbb{R}^d)}\leq ct^{-(d-1)(\frac{1}{2}-\frac{1}{q})}\|\psi\|_{\dot{H}^{\frac{d-1}{2}-\frac{d+1}{q},p}(\mathbb{R}^d)}.\label{04091}
\end{align}

(ii) For any $\varphi\in L^s(\mathbb{R}^d)$,
\begin{align}
\|K(t)\varphi\|_{L^r(\mathbb{R}^d)}\leq C|t|^{1-\delta(r)+\delta(s)}\|\varphi\|_{L^s(\mathbb{R}^d)}\label{7283}
\end{align}
with
\begin{equation}
\label{7284}\left\{
\begin{array}{lll}
0\leq \delta(r)-\delta(s)\leq \min\{1+\gamma(r), d(1-\gamma(r))\},\\
1<s, r<\infty\quad {\rm if}\quad d=2,
\end{array}\right.
\end{equation}
}

A fact on the estimates for the operator $K(t)$ acting in homogeneous Besov spaces below(see Lemma 2.2 in \cite{Ginibre1989}).

{\bf Lemma 2.2.} {\it (i) For $2\leq r\leq \infty$ and $\frac{1}{r}+\frac{1}{\bar{r}}=1$,
\begin{align}
\|K(t)\varphi; \dot{B}^{\rho}_r\|\leq C|t|^{-\gamma(r)}\|\varphi;\dot{B}^{\rho+2\eta(r)-1}_{\bar{r}}\|,\label{4181}
\end{align}
where $\|\varphi;\dot{B}^{\rho+2\eta(r)-1}_{\bar{r}}\|$ is the norm of $\varphi$ in Besov space.

(ii) For $d\geq 2$, $0\leq \gamma(r)\leq 1$,
\begin{align}
\|K(t)\varphi; \dot{B}^{\rho}_r\|\leq C|t|^{-\mu}\|\varphi;\dot{B}^{\rho_1}_{r_1}\|\label{7301}
\end{align}
for all $\rho, \rho_1, r_1, \mu$ satisfying
\begin{align}
0&\leq 1+\mu=\rho+\delta(r)-\rho_1-\delta(r_1)\nonumber\\
&\leq \frac{1}{2}[\gamma(r)-\gamma(r_1)][1+\frac{1}{\gamma(r)}]\leq 1+\gamma(r).\label{7302}
\end{align}

}

We will recall some results on harmonic analysis below.

{\bf Definition 2.3.} A pair $(q,r)$ of positive real numbers is said to be wave admissible if $2\leq q\leq +\infty$, $2\leq r<+\infty$ and
$$
\frac{1}{q}\leq \frac{d-1}{2}\left(\frac{1}{2}-\frac{1}{r}\right).
$$

{\bf Lemma 2.4.(Strichartz estimate, \cite{Keel})} {\it Take two admissible pairs $(q_1,r_1)$ and $(q_2,r_2)$, let $\frac{1}{q_2}+\frac{1}{\bar{q}_2}=1$, $\frac{1}{r_2}+\frac{1}{\bar{r}_2}=1$,
\begin{align}
\frac{1}{q_1}+\frac{d}{r_1}=\frac{d}{2}-s=\frac{1}{\bar{q}_2}+\frac{d}{\bar{r}_2}-2.\label{418w1}
\end{align}
Then for any $I\subset \mathbb{R}$,
\begin{align}
&\quad \|(u,\partial_t u)\|_{C_t(I;\dot{H}^s\oplus\dot{H}^{s-1})}+\|u\|_{L^{q_1}_tL^{r_1}_x(I\times \mathbb{R}^d)}\nonumber\\
&\leq \|(u(0),\partial_tu(0))\|_{\dot{H}^s\oplus\dot{H}^{s-1}}+\||\nabla|(\partial^2_tu-\Delta u)\|_{L^{\bar{q}_2}_tL^{\bar{r}_2}_x(I\times \mathbb{R}^d)}.\label{418w2}
\end{align}
}

The next lemmas are about product rule and fractional chain rules, we can see \cite{Bourgain, Kenig1993, Li} for more details.

 {\bf Lemma 2.5.} {\it Assume that $s\in (0,1]$, $1<r, r_1,r_2,q_1,q_2<+\infty$ and satisfy $\frac{1}{r}=\frac{1}{r_1}+\frac{1}{q_1}=\frac{1}{r_2}+\frac{1}{q_2}$. Then
\begin{align*}
\||\nabla|^s(fg)\|_{L^r_x}\lesssim \|f\|_{L^{r_1}_x}\||\nabla|^sg\|_{L^{q_1}_x}+\||\nabla|^sf\|_{L^{r_2}_x}\|g\|_{L^{q_2}_x}.
\end{align*}
}

{\bf Lemma 2.6.} {\it Assume that $s\in (0,1]$, $1<q,q_1,q_2<+\infty$ and satisfy $\frac{1}{q}=\frac{1}{q_1}+\frac{1}{q_2}$, $G\in C^1(\mathbb{C})$. Then
\begin{align*}
\||\nabla|^sG(u)\|_{L^q_x}\lesssim \|G'(u)\|_{L^{q_1}_x}\||\nabla|^su\|_{L^{q_2}_x}.
\end{align*}
}

{\bf Lemma 2.7.} {\it
Assume that $G$ is a H\"{o}lder continuous function of order $0<p<1$. Then
\begin{align*}
\||\nabla|^sG(u)\|_{L^q_x}\lesssim \||u|^{p-\frac{s}{\sigma}}\|_{L^{q_1}_x}\||\nabla|^{\sigma}u\|_{L^{\frac{sq_2}{\sigma}}_x}^{\frac{s}{\sigma}}
\end{align*}
for every $0<s<p$, $1<q<\infty$ and $\frac{s}{p}<\sigma<1$ satisfying $\frac{1}{q}=\frac{1}{q_1}+\frac{1}{q_2}$ and $(1-\frac{s}{p\sigma})q_1>1$.
}

{\bf Lemma 2.8.} {\it
Assume that $G$ is a H\"{o}lder continuous function of order $0<p\leq 1$, $0<s<\sigma p<p$, $1<q,q_1,q_2,r_1,r_2,r_3<\infty$ and satisfy
\begin{align*}
(1-p)r_1>1,\quad (p-\frac{s}{\sigma})r_2>1,\\
\frac{1}{q}=\frac{1}{q_1}+\frac{1}{q_2}=\frac{1}{r_1}+\frac{1}{r_2}+\frac{1}{r_3}.
\end{align*}
 Then
\begin{align*}
&\quad\||\nabla|^s[w\cdot(G(u+v)-G(u))]\|_{L^q_x}\\
&\lesssim\||\nabla|^{\sigma}w\|_{L^{q_1}_x}\|v\|_{L^{pq_2}_x}^p
+\|w\|_{L^{r_1}}
\|v\|_{L^{(p-\frac{s}{\sigma})r_2}}^{p-\frac{s}{\sigma}}
\left(\||\nabla|^{\sigma}u\|_{L^{\frac{sr_3}{\sigma}}_x}+\||\nabla|^{\sigma}v\|_{L^{\frac{sr_3}{\sigma}}_x}\right)^{\frac{s}{\sigma}}.
\end{align*}
}

\section{Local wellposedness results on (\ref{1})}
\qquad In this section, we will consider the global wellposedness of $H^1\oplus H^1$ solution to (\ref{1}).

\subsection{$H^1\oplus H^1$ global wellposedness results when $d=2$ and $d=3$}
\qquad
Let $d=2$ or $d=3$,
\begin{equation}
\label{7261}\left\{
\begin{array}{lll}
1\leq l,r, q, q_1\leq \infty,\quad \frac{2\eta(r)}{(d+1)}=\frac{\gamma(r)}{(d-1)}=\frac{\delta(r)}{d}=\frac{1}{2}-\frac{1}{r},\\
1<r<\infty \ {\rm if} \ d=2,\quad |\gamma(r)|\leq 1 \ {\rm if} \ d=3,\\
\frac{(\alpha+\beta+2)d}{l}\leq \min \{1+\gamma(r), d(1-\gamma(r))\},\\
\frac{(\alpha+\beta+2)}{q}+\frac{1}{q_1}\leq 1,\\
\eta_1=2-(\alpha+\beta+2)(\frac{d}{l}+\frac{1}{q})>0.
\end{array}\right.
\end{equation}
And for any interval $I$ containing $0$, denote
\begin{align}
X_0(I)=L^q(I, L^l), \quad X_1(I)=L^{q_1}(I, L^r).\label{726w1}
\end{align}

The following lemma shows some form of equivalence between the system of partial differential equations (\ref{1}) and the system of integral equations (\ref{771}) and gives the uniqueness of the solution to (\ref{771}).

{\bf Lemma 3.1.} {\it  Let $d\geq 2$, $I$ be a bounded interval containing $0$, $u_{10}\in X_0(I)\cap X_1(I)$ and $v_{10}\in X_0(I)\cap X_1(I)$. Then

(1) For any $t_0\in I$, $F_1(t_0; u, v)$ and $F_2(t_0; u, v)\in X_1(I)$ are continuous functions of $t_0$ with values $F_1(t_0; u,v)\in X_1(I)$ and $F_2(t_0; u,v)\in X_1(I)$.
And for any $(u_1,v_1), (u_2,v_2)\in [X_1(I)\cap X_0(I)]^2$, there holds
\begin{align}
&\quad \|F_1(t_0;u_1,v_1)-F_1(t_0;u_2,v_2);X_1(I)\|+\|F_2(t_0;u_1,v_1)-F_2(t_0;u_2,v_2);X_1(I)\|\nonumber\\
&\leq C[\|u_1-u_2;X_1(I)\|+\|v_1-v_2;X_1(I)\|]\nonumber\\
&\qquad \times\left\{|I|^2+|I|^{\eta_1}\left(\sum_{i=1}^2[\|u_i;X_0(I)\|^{\alpha+\beta+2}+\|v_i;X_0(I)\|^{\alpha+\beta+2}]\right)\right\}.\label{7262}
\end{align}

(2) For any $t_1,t_2\in I$, $G_1(t_1,t_2;u,v)$ and $G_2(t_1,t_2;u,v)$ are continuous functions of $t_1,t_2$ and $G_1(t_1,t_2;u,v)\in X_{1loc}(\mathbb{R})$, $G_2(t_1,t_2;u,v)\in X_{1loc}(\mathbb{R})$. For any bounded interval $J\supset I$ and $(u_1,v_1), (u_2,v_2)\in [X_1(I)\cap X_0(I)]^2$, there holds
\begin{align}
&\quad \|F_1(t_1,t_2;u_1,v_1)-F_1(t_1,t_2;u_2,v_2);X_1(J)\|+\|F_2(t_1,t_2;u_1,v_1)-F_2(t_1,t_2;u_2,v_2);X_1(J)\|\nonumber\\
&\leq C[\|u_1-u_2;X_1([t_1,t_2])\|+\|v_1-v_2;X_1([t_1,t_2])\|]\nonumber\\
&\qquad \times\left\{|J|^2+|J|^{\eta_1}\left(\sum_{i=1}^2[\|u_i;X_0([t_1,t_2])\|^{\alpha+\beta+2}+\|v_i;X_0([t_1,t_2])\|^{\alpha+\beta+2}]\right)\right\}.\label{7282}
\end{align}

(3) For any $t_0\in I$, $F_1(t_0;u,v)$ and $F_2(t_0;u,v)$ satisfy $\square F_1(t_0;u,v)=\lambda|u|^{\alpha}|v|^{\beta+2}u$ and $\square F_2(t_0;u,v)=\mu|u|^{\alpha+2}|v|^{\beta}v$ in $\mathfrak{D}'(I\times \mathbb{R}^d)$ respectively,  and for any $t_1,t_2\in I$, $\square G_1(t_1,t_2;u,v)=0$ and $\square G_2(t_1,t_2;u,v)=0$ in $\mathfrak{D}'(\mathbb{R}^{d+1})$.

(4) Let $u_{10}, v_{10}\in X_{1loc}(I)$. Then the system (\ref{771}) has at most one solution in $X_{1loc}(I)\cap X_{2loc}(I)$.
}

{\bf Proof:} (1) and (2)

 By (\ref{7283}) and (\ref{7284}), using H\"{o}lder inequality, we have
\begin{align}
&\quad\|K(t-\tau)[f_1(u_1(\tau),v_1(\tau))-f_1(u_2(\tau),v_2(\tau))]\|_r+\|K(t-\tau)[f_2(u_1(\tau),v_1(\tau))-f_2(u_2(\tau),v_2(\tau))]\|_r\nonumber\\
&\leq C|t-s|^{1-\delta(r)+\delta(s)}(\|[f_1(u_1(\tau),v_1(\tau))-f_1(u_2(\tau),v_2(\tau))]\|_s+\|[f_2(u_1(\tau),v_1(\tau))-f_2(u_2(\tau),v_2(\tau))]\|_s)\nonumber\\
&\leq C|t-s|^{1-\delta(r)+\delta(s)}[\|u_1(\tau)-u_2(\tau)\|_r+\|v_1(\tau)-v_2(\tau)\|_r]\nonumber\\
&\qquad \times \{\|u_1(\tau)\|^{\alpha+\beta+2}_l+\|u_2(\tau)\|^{\alpha+\beta+2}_l+\|v_1(\tau)\|^{\alpha+\beta+2}_l+\|v_2(\tau)\|^{\alpha+\beta+2}_l\}\label{7285}
\end{align}
with
\begin{align}
\delta(r)-\delta(s)=\frac{(\alpha+\beta+2)d}{l}.\label{7286}
\end{align}
By the expressions of $F_1(u,v)$, $F_2(u,v)$, $G_1(u,v)$ and $G_2(u,v)$, using (\ref{7285}) and Young's inequality for the integral, we can obtain
(\ref{7262}) and (\ref{7282}).

(3) can be obtained by the standard duality argument and some elementary computations, we omit the details here.

(4) Assume that $(u_1,v_1)$ and $(u_2,v_2)$ be two solutions of (\ref{771}) with the same initial data $(u_0,v_0)$.
Then
\begin{align*}
u_1-u_2=F_1(u_1,v_1)-F_1(u_2,v_2),\quad v_1-v_2=F_2(u_1,v_1)-F_2(u_2,v_2).
\end{align*}
Replacing $I$ in (\ref{7262}) by a sufficient small interval $J$ containing $0$ such that
\begin{align*}
C\left\{|J|^2+|J|^{\eta_1}\left(\sum_{i=1}^2[\|u_i;X_0(J)\|^{\alpha+\beta+2}+\|v_i;X_0(J)\|^{\alpha+\beta+2}]\right)\right\}\leq \frac{1}{2},
\end{align*}
we can get $$
[\|u_1-u_2;X_1(J)\|+\|v_1-v_2;X_1(J)\|]\leq \frac{1}{2}[\|u_1-u_2;X_1(J)\|+\|v_1-v_2;X_1(J)\|],
$$
which means that $(u_1,v_1)=(u_2,v_2)$ in $J$. Iterating the process, we know that  $(u_1,v_1)=(u_2,v_2)$ everywhere in $I$.\hfill$\Box$

For any interval $I\subset \mathbb{R}$ and for suitable values of $\rho$, $r$ and $q$, we will introduce the notations $X_2(I)=L^q(I,\dot{B}^{\rho}_r)$ and let $B_i(I,R)$ be the closed ball of radius $R$ in $X_i(I)$, $i=1,2$. We have the following estimate

{\bf Lemma 3.2.} {\it Let $d\geq 2$, $\rho$ and $r$ satisfy $q\geq (\alpha+\beta+3)$ and $0\leq \rho<1$,
\begin{align}
&0\leq \gamma(r)\leq \frac{(d-1)}{(d+1)},\quad (\alpha+\beta+2)(\frac{d}{r}-\rho)\leq 1+\gamma(r),\label{7303}\\
&\eta_2=2-(\alpha+\beta+2)(\frac{d}{r}-\rho+\frac{1}{q})>0.\label{7304}
\end{align}
Let $I$ be a bounded open interval containing $0$ and $u,v\in X_2(I)$. Then

(1) $F_1(u,v)\in X_2(I)$, $F_2(u,v)\in X_2(I)$, and there holds
\begin{align}
&\quad \|F_1(u,v);X_2(I)\|+\|F_2(u,v);X_2(I)\|\nonumber\\
&\leq C\{|I|^2(\|u;X_2(I)\|+\|v;X_2(I)\|)+|I|^{\eta_2}(\|u;X_2(I)\|^{\alpha+\beta+3}+\|v;X_2(I)\|^{\alpha+\beta+3})\}.\label{7305}
\end{align}

(2) For any bounded interval $J\supset I$ and for any $t_1,t_2\in I$, $G_1(t_1,t_2;u,v)$ and $G_2(t_1,t_2; u,v)$ are continuous functions of $t_1, t_2$ with values in $X_2(J)$, and there holds
\begin{align}
&\quad \|G_1(t_1,t_2;u,v); X_2(J)\|+\|G_2(t_1,t_2;u,v); X_2(J)\|\nonumber\\
&\leq C\{|J|^2(\|u;X_2([t_1,t_2])\|+\|v;X_2([t_1,t_2])\|)\nonumber\\
&\quad+ |J|^{\eta_2}(\|u;X_2([t_1,t_2])\|^{\alpha+\beta+3}+\|v;X_2([t_1,t_2])\|^{\alpha+\beta+3})\}.\label{7306}
\end{align}
}

{\bf Proof:} First we prove that
\begin{align}
&\quad \|f_1(u,v);\dot{B}^{\theta}_{l,m}\|+\|f_2(u,v);\dot{B}^{\theta}_{l,m}\|\nonumber\\
&\leq C(\|u;\dot{B}^{\theta}_{k,m}\|+\|v;\dot{B}^{\theta}_{k,m}\|)(\||u|^{\alpha+\beta+2}\|_s+\||v|^{\alpha+\beta+2}\|_s)\label{7311}
\end{align}
for $0<\theta<1$, $1\leq l\leq k\leq \infty$, $1\leq m\leq \infty$ and $\frac{1}{s}=\frac{1}{l}-\frac{1}{k}$. Here $\dot{B}^{\theta}_{l,m}$ was defined as that in the Appendix of \cite{Ginibre1985}
\begin{align}
&\dot{B}^{\theta}_{l,m}=\{w\in \mathfrak{L}':\{\sum_j2^{jm\theta}\|\varphi_j*w\|_l^m\}^{\frac{1}{m}}\equiv \|w;\dot{B}^{\theta}_{l,m}\|<\infty\},\label{7312}\\
&\mathfrak{L}=\{\mathcal{\varphi}: D^{\alpha} \hat{u}(0)=0\quad {\rm for \ any \ multindex}\ \alpha \}. \label{10111}
\end{align}
Here $\mathcal{\varphi}$ is the completion of $\mathcal{S}(\mathbb{R}^d)$, $\mathcal{S}(\mathbb{R}^d)$ is the set of all complex-valued rapidly decreasing infinitely differentiable functions defined on the d-dimension real Euclidean space $\mathbb{R}^d$.

Let $\tau_y$ be the space translation by $y\in \mathbb{R}^d$. Then
\begin{align}
|\tau_y f_1(u,v)-f_1(u,v)|&\leq |\tau_y u-u|\int^1_0[|\frac{\partial f_1}{\partial u}((\vartheta_1\tau_y u+(1-\vartheta_1)u),\tau_yv)|]d\vartheta_1\nonumber\\
&\quad+|\tau_y v-v|\int^1_0[|\frac{\partial f_1}{\partial v}(u, (\vartheta_2\tau_y v+(1-\vartheta_2)v))|]d\vartheta_2,\label{7313}\\
|\tau_y f_2(u,v)-f_2(u,v)|&\leq |\tau_y u-u|\int^1_0[|\frac{\partial f_2}{\partial u}((\vartheta_3\tau_y u+(1-\vartheta_3)u),\tau_yv)|]d\vartheta_3\nonumber\\
&\quad+|\tau_y v-v|\int^1_0[|\frac{\partial f_2}{\partial v}(u, (\vartheta_4\tau_y v+(1-\vartheta_4)v))|]d\vartheta_4.\label{731w1}
\end{align}
Applying H\"{o}lder's inequality to (\ref{7313}) and (\ref{731w1}), we can get
\begin{align}
&\quad\|\tau_y f_1(u,v)-f_1(u,v)\|_l+\|\tau_y f_2(u,v)-f_2(u,v)\|_l\nonumber\\
&\leq C_0[\|\tau_y u-u\|_k+\|\tau_y v-v\|_k][\||u|^{\alpha+\beta+2}\|_s+\||v|^{\alpha+\beta+2}\|_s],\label{731w2}
\end{align}
which can deduce (\ref{7311}) immediately.

By (\ref{7311}) and the Sobolev inequality, we can obtain
\begin{align}
&\quad \|K(t-\tau)f_1(u,v);\dot{B}^{\rho}_r\|+\|K(t-\tau)f_2(u,v);\dot{B}^{\rho}_r\|\nonumber\\
&\leq C|t-\tau|^{-\mu}[\|u;\dot{B}^{\rho'}_r\|^{\alpha+\beta+3}+\|v;\dot{B}^{\rho'}_r\|^{\alpha+\beta+3}]\label{731w3}
\end{align}
for $\rho'\leq \rho$ and
$$(\alpha+\beta+3)(\frac{d}{r}-\rho)=\frac{d}{r'}-\rho'=\rho+\delta(r)-\rho'-\delta(r')=1+\mu.$$

For given $r$ in range (\ref{7303}), if (\ref{7304}) holds, then $r'$ satisfies (\ref{7302}) and (\ref{7301}) holds for these $\rho, r, \rho', r'$ and $\mu$.
Applying (\ref{7301}) to $f_1(u,v)$ and $f_2(u,v)$, then estimating the time integral by Young's inequality, we can obtain (\ref{7305}) and (\ref{7306}).\hfill $\Box$

Now we will prove the local existence and uniqueness of the solution to (\ref{771}).

{\bf Proposition 3.1.} {\it Let $d\geq 2$, $\rho, r, q$ and $q_1$ satisfy $0\leq \rho<1$, $1\leq q\leq q_1\leq \infty$, (\ref{7303}) and (\ref{7304}) hold. Then for any
$R>0$, there exists $T(R)>0$ such that for any $u_{10}, v_{10}\in B_2(I,R)\cap X_1(I)$ with $I=[-T(R), T(R)]$, (\ref{771}) possesses a solution $(u,v)$ satisfying
$u,v\in B_2(I,2R)\cap X_1(I)$ and
$$
[\|u;X_1(I)\|+\|v;X_1(I)\|]\leq 2[\|u_{10};X_1(I)\|+\|v_{10};X_1(I)\|].
$$
 And it is unique in $X_2(I)\cap X_1(I)$. }

{\bf Proof:} Let $\frac{d}{l}=\frac{d}{r}-\rho$ in Lemma 3.1 and Lemma 3.2 such that (\ref{7261}), (\ref{7303}) and (\ref{7304}) hold with $\eta_1=\eta_2=\eta$.
Meanwhile, using Sobolev embedding, we have
\begin{align*}
& \dot{B}^{\rho}_r\subset L^l,\quad \|u\|_l\leq \bar{C}\|u;\dot{B}^{\rho}_r\|,\quad \|v\|_l\leq \bar{C}\|v;\dot{B}^{\rho}_r\|,\\
& X_2(\cdot)\subset X_0(\cdot),\quad \|u;X_0\|\leq \bar{C}\|u;X_2\|,\quad \|u;X_0\|\leq \bar{C}\|u;X_2\|.
\end{align*}
If we choose $T=T(R)$ small enough such that
\begin{align}
 C\{4T^2+(2T)^{\eta}(2R)^{\alpha+\beta+2}(1+2\bar{C}^{\alpha+\beta+2})\}\leq \frac{1}{2},\label{8011}
\end{align}
then apply the estimates of Lemma 3.1 and Lemma 3.2 to the operators $A_1(u_{10},u_{20})$ and $A_2(v_{10},v_{20})$, we find that they make the set $S=B_2(I,2R)\cap X_1(I)$ invariant and they are contracting on $S$ in the norm of $X_1(I)$. Since $B_1(I,R_1)\cap B_2(I,2R)$ is $w^*$-compact in $X_1(I)\cap X_2(I)$ for any $R_1>0$, especially, it is compact in the $w^*$-topology of $X_1(I)$, therefore, it is $w^*$-closed and strongly closed in $X_1(I)$ and consequently $S$ is strongly closed in $X_1(I)$. By the contraction mapping theorem, we obtain the existence and uniqueness of a solution to (\ref{771}) in $S$. While the uniqueness result is the directly following from Lemma 3.1 (3).\hfill $\Box$

To establish the existence of the solution to (\ref{771}) corresponding to initial data with finite energy, we will recall some facts below.

{\bf Fact 1:} Let $d\geq 2$,
\begin{align}
X_e=\{(\varphi,\psi):\varphi\in H^1, \psi\in L^2\}=H^1\oplus L^2,\label{801x1}
\end{align}
and $(\varphi,\psi)\in X_e$ is associated a solution of the free equation
\begin{align}
\varphi(t)=\dot{K}(t)\varphi+K(t)\psi\label{801x2}
\end{align}
in $L(\mathbb{R},H^1)$. Then for any $(\varphi,\psi)\in X_e$, $\varphi\in X_2(\mathbb{R})$ and satisfies
\begin{align}
\|\varphi;X_2(\mathbb{R})\|\leq C(\|\nabla \varphi\|_2+\|\psi\|_2).\label{801x3}
\end{align}
Here $X_2(\mathbb{R})=L^q(\mathbb{R},\dot{B}^{\rho}_r)$ and $\rho, r, q$ satisfy
\begin{equation}
\label{801x4}\left\{
\begin{array}{lll}
&0\leq \delta(r)\leq \frac{d}{2},\quad -1\leq \sigma=\rho+\delta(r)-1<\frac{1}{2},\\
&\sigma\leq \frac{\gamma(r)}{2},\quad \frac{1}{q}=\max(0,\sigma).
\end{array}\right.
\end{equation}

{\bf Fact 2:} Let $d\geq 2$. There exist $\rho, r$ and $q$ such that $0\leq \rho<1$, (\ref{7303}), (\ref{7304}) and (\ref{801x4}) hold.

By Fact 1, Fact 2,  using Lemma 3.1 with $\frac{d}{l}=\frac{d}{r}-\rho$ and Lemma 3.2, we can obtain the results on the existence of the solution to (\ref{771}) corresponding to initial data with finite energy as follows.

{\bf Proposition 3.2.} {\it Let $d\geq 2$. $X_1$ and $X_2$ correspond to values of $\rho, r,q$ provided by Fact 2 and $q_1\geq q$. Then

(1) For any $(u_{10},u_{20})\in X_e$ and $(v_{10},v_{20})\in X_e$, there exists $T>0$ depending only on $\|(u_{10},u_{20}); X_e\|$ and $\|(v_{10},v_{20}); X_e\|$ such that (\ref{771}) has a unique solution in $X_1(I)\cap X_2(I)$ with $I=[-T,T]$.

(2) For any $(u_{10},u_{20})\in X_e$ and $(v_{10},v_{20})\in X_e$, and any interval $I$, (\ref{771}) has at most one solution $(u,v)$ with $u,v\in X_{1loc}(I)\cap X_{2loc}(I)$.
}

For $(u,v)\in X_1(I)\cap X_2(I)$, we define the weighted energy
\begin{align}
E_w(u,v)=\mu(\|\nabla u\|^2_2+\|u_t\|^2_2)+\lambda(\|\nabla v\|^2_2+\|v_t\|^2_2)+\lambda\mu\int_{\mathbb{R}^d}|u|^{\alpha+2}|v|^{\beta+2}dx.\label{802x1}
\end{align}
Choosing an even nonnegative function $h_1\in \mathfrak{L}^{\infty}(\mathbb{R}^d)$ with compact support and such that $\|h_1\|_1=1$, we can define $h_j(x)=j^dh_1(jx)$,
\begin{align*}
f_{1j}(u,v)=h_j*f_1(h_j*u,h_j*v),\quad f_{2j}(u,v)=h_j*f_2(h_j*u,h_j*v),
\end{align*}
and
\begin{align}
E_{jw}(u,v)=\mu(\|\nabla u\|^2_2+\|u_t\|^2_2)+\lambda(\|\nabla v\|^2_2+\|v_t\|^2_2)+\lambda\mu\int_{\mathbb{R}^d}|h_j*u|^{\alpha+2}|h_j*v|^{\beta+2}dx.\label{802x1}
\end{align}
Consider the regularized system
\begin{align}
u=h_j*u^{(0)}+F_{1j}(u,v),\quad v=h_j*v^{(0)}+F_{2j}(u,v),\label{802w1}
\end{align}
where $F_{1j}$ and $F_{2j}$ are defined by (\ref{771}) with $f_1, f_2$ replaced by $f_{1j}, f_{2j}$.

Noticing the fact that the convolution with $h_j$ is a contraction in the spaces $L^r$ and $\dot{B}^{\rho}_r$, all the estimates above hold with $f$ replaced by $f_j$, we know that Proposition 3.2, part (1) holds for the system (\ref{802w1}) with the same $T$, independently of $j$, and the solution $(u_j,v_j)$ to (\ref{802w1}) obtained in part (1) converges to the solution of (\ref{771}) in $X_1(I)$ when $j\rightarrow \infty$. Moreover, the following properties hold

(1) For any nonnegative integer $k$, $(u_j, \dot{u}_{jt}), (v_j, \dot{v}_{jt})\in \mathfrak{L}^1(I, H^{k+1}\oplus H^k$ and $(u_j,v_j)$ satisfies the system
\begin{align*}
\square u_j=f_{1j}(u_j,v_j),\quad \square v_j=f_{2j}(u_j,v_j).
\end{align*}

(2)
\begin{align}
E_{jw}(u_j(t), v_j(t))=E_{jw}(h_j*u_0, h_j*v_0)\equiv E_{jw}\label{802w2}
\end{align}
and the estimates
\begin{align}
&\|u_j(t)\|_2+\|v_j(t)\|_2\leq Ce(E_{jw},t)\leq Ce(\bar{E}_w,t),\label{802w3}\\
&\|\nabla u_j(t)\|^2_2+\|u_{jt}(t)\|^2_2+\|\nabla v_j(t)\|^2_2+\|v_{jt}(t)\|^2_2\leq C[\dot{e}_t(E_{jw},t)]^2\leq C[\dot{e}_t(\bar{E}_w,t)]^2\label{802w4}
\end{align}
with
\begin{align}
\bar{E}_w=\sup E_{jw}<\infty\label{8041}
\end{align}
and
\begin{align}
e(E_w,\tau)&=\|u_0\|_2 \cosh (a|\tau|)+(E_w+a^2\|u_0\|_2^2)^{\frac{1}{2}}a^{-1}\sinh (a|\tau|)\nonumber\\
&\quad +\|v_0\|_2 \cosh (a|\tau|)+(E_w+a^2\|v_0\|_2^2)^{\frac{1}{2}}a^{-1}\sinh (a|\tau|).\label{8042}
\end{align}
Particularly, $(u_j,\dot{u}_{jt})$ and $(v_j,\dot{v}_{jt})$ are locally bounded in $H^1\oplus L^2$ uniformly in $j$.

The global existence and uniqueness results for finite energy solutions of (\ref{771}) are based on the following conservation of energy for
these solutions with finite energy initial data.

{\bf Proposition 3.3.} {\it  Let $d\geq 2$, $(u_{10}, u_{20}), (v_{10},v_{20})\in X_e$ and $I$ be an open interval containing $0$. Let $\rho, r$ and $q$ satisfy $0\leq \rho<1$, (\ref{7303}), (\ref{7304}), (\ref{801x4}) and $q_1=\infty$. Assume that $(u^{(0)}, v^{(0)})$ satisfies (\ref{801x2}) and $(u,v)$ is a solution of (\ref{771}) in $X_1(I)\cap X_2(I)$. Then $(u,u_t), (v,v_t)\in \mathfrak{L}(I, H^1\oplus L^2$ and $(u,v)$ satisfies the conservation of energy
\begin{align}
E_w(u(t),v(t))=E_w(u_0,v_0)\equiv E_w\label{804x1}
\end{align}
and
\begin{align}
&\|u(t)\|_2+\|v(t)\|_2\leq Ce(E_w,t),\label{804x2}\\
&\|\nabla u(t)\|^2_2+\|u_t(t)\|^2_2+\|\nabla v(t)\|^2_2+\|v_t(t)\|^2_2\leq C[\dot{e}(E_w,t)]^2\label{804x3}
\end{align}
for all $t\in I$.
}

{\bf Proof:} We just need to prove the result in any bounded subinterval $I'\subset\subset I$ containing $0$.

Let
\begin{align}
R=\sup_{s\in I'}[\|u^{(0)}+G_1(t,u,v);X_2(I')\|+\|v^{(0)}+G_2(t,u,v);X_2(I')\|].\label{804w1}
\end{align}
By (\ref{7282}), $R$ is finite. Let $T=T(R)$ be defined as in the proof of Proposition 2.1. By the results of Proposition 2.1, for any $t\in I'$, solving (\ref{771})
with initial time $t$ by contraction, the solution $(u,v)$ can be recovered in the interval $I'\cap [t-T,t+T]$. If we cover $I'$ by a finite number of intervals $I_k$ of length $2T$ and centers at $t_k=(1-\epsilon) kT$ for some $\epsilon>0$, then we can deduce the result in $I'$ by the corresponding one in $I_k$ for successive values of $k=0$, $\pm 1$, $\pm 2$,.... Therefore, we only need to prove this proposition in the special case where $I$ is a small interval containing $0$ where (\ref{771}) can be solved by the contraction method of Proposition 2.1 below.

Approximating $(u,v)$ in $I$ by the regularized solutions $(u_j,v_j)$ of (\ref{802w1}), using (\ref{802w4}), (\ref{8042}) and standard compactness arguments, it follows that
$(u_j,v_j)$ converges to $(u,v)$ in the $w^*$-sense in $L^{\infty}(I,H^1)$  and $(u_{jt},v_{jt})$ converges to $(u_t,v_t)$ in the $w^*$-sense in $L^{\infty}(I,L^2)$. Moreover, $(u,u_t)$ and $(v,v_t)$ satisfy (\ref{804x2}) and (\ref{804x3}) for almost all $t$ in $I$. Here $u_t$ and $v_t$ are the derivatives of $u$ and $v$ in $\mathcal{D}'(I,L^2)$. Hence $(u,v)\in \mathcal{C}(I, L^s)$ because $u,v\in \mathcal{C}(I, L^r)\cap L^{\infty}(I,H^1)$ and $u_t,v_t\in L^{\infty}(I,L^2)$. By Corollary 2.1, $(u,v)$ satisfies (\ref{1}) in $\mathcal{D}'(I\times \mathbb{R}^d)$ and $u_{tt},v_{tt}\in L^{\infty}(I,H^{-1})$. Consequently, $u_t,v_t\in \mathcal{C}(I,H^{-1})$ and
$u_t,v_t\in \mathcal{C}_w(I,L^2)$. (\ref{804x2}) and (\ref{804x3}) hold for all $t\in I$ by the continuity properties of $(u,u_t)$ and $(v, v_t)$.

By uniform boundedness in $H^1$ and convergence in $\mathcal{C}(I,L^r)$, using interpolation, we know that $u_j$ and $v_j$ respectively converge to $u$ and $v$ in $\mathcal{C}(I,L^s)$ for all $s$ satisfying $2<s<\frac{2d}{d-2}$. Similar to (3.63) and (3.64) in \cite{Ginibre1985}, it is easy to show that $u_j$ and $v_j$ respectively converge to $u$ and $v$ in $\mathcal{C}(I,L^2)$, $u_{jt}$ and $v_{jt}$ respectively converge to $u_t$ and $v_t$ weakly in $L^2$ for each $t\in I$. Consequently, $u_j(t)$ and $v_j(t)$ respectively converge to $u(t)$ and $v(t)$ weakly in $H^1$ for each $t\in I$.

Letting the limit $j\rightarrow \infty$ in (\ref{802w2}), by the convergence of $u_j, v_j$ to $u,v$ in $\mathcal{C}(I,L^s)$ for $2\leq s\leq \alpha+\beta+4$,
we can directly show that
$$
\int_{\mathbb{R}^3}|h_j*u|^{\alpha+2}|h_j*v|^{\beta+2}dx\rightarrow \int_{\mathbb{R}^3}|u|^{\alpha+2}|v|^{\beta+2}dx,\quad E_{jw}(u(t),v(t))\rightarrow E_w(u_0,v_0).
$$
Since $(u_j(t),u_{jt}(t))$ and $(v_j(t),v_{jt}(t))$ converge to $(u(t),u_t(t))$ and $(v(t),v_t(t))$ in $H^1\oplus L^2$, we get
\begin{align}
E_w(u(t),v(t))\leq E_w(u_0,v_0)\quad {\rm for \ all}\quad t\in I.\label{8051}
\end{align}
The time reversal invariance of (\ref{1}) and the results of Proposition 2.1 imply (\ref{804x1}) for all $t\in I$. Since $u,v\in \mathcal{C}(I,L^s)$ for $2\leq s\leq \alpha+\beta+4$, we know that $\|u(t)\|^2_2+\|\nabla u(t)\|^2_2+\|u_t(t)\|^2_2$ and $\|v(t)\|^2_2+\|\nabla v(t)\|^2_2+\|v_t(t)\|^2_2$ are continuous functions of time, and the weak continuity implies strong continuity of $(u,u_t)$ as well as $(v,v_t)$ in $H^1\oplus L^2$ as a function of time.

Similar to the arguments above, it is easy to show that $(u_j(t),u_{jt}(t))$ and $(v_j(t),v_{jt}(t))$ respectively converge to $(u(t),u_t(t))$ and $(v(t),v_t(t))$ in $H^1\oplus L^2$ for each $t\in I$. \hfill $\Box$

Now we can establish the global existence and uniqueness result for finite energy solutions.

{\bf Theorem 3.1.} {\it Let $d=2$ or $d=3$, $(u_{10},u_{20})\in X_e$ and $(v_{10},v_{20})\in X_e$. Then (\ref{771}) with (\ref{801x2}) has a unique solution $(u,v)$ such that $(u,u_t)\in \mathcal{C}(\mathbb{R},X_e)$, $(v,v_t)\in \mathcal{C}(\mathbb{R},X_e)$ and (\ref{804x1})--(\ref{804x3}) hold.

Suppose that $\rho,r,q$ and $q_1$ satisfy $0\leq \rho<1$, $q_1\geq q$, (\ref{7303}), (\ref{7304}) and (\ref{801x4}). Then the solution $(u,v)$ is unique and $u,v\in X_{1loc}(\mathbb{R})\cap X_{2loc}(\mathbb{R})$. }

{\bf Proof:} Under the assumptions on $\rho, r, q$ and $q_1$, by Proposition 3.2, (\ref{771}) with initial data $(u_{10}, u_{20}), (v_{10},v_{20})\in X_e$ can be solved locally in time with $u(t), v(t)\in X_1(I)\cap X_2(I)$. Here the length of the interval $I$ depends on the norms of $(u_{10}, u_{20})$ and $(v_{10},v_{20})$ in $X_e$.
By Proposition 3.3, the local solutions thereby obtained satisfy $u(t),v(t)\in \mathcal{C}(I,X_e)$, (\ref{804x1}), (\ref{804x2}) and (\ref{804x3}), which imply the global existence of the solution $(u(t),v(t))$ with $u(t), v(t)\in X_{1loc}(\mathbb{R})\cap X_{2loc}(\mathbb{R})\cap \mathcal{C}(\mathbb{R},X_e)$.

Note that a function in $L^{\infty}_{loc}(\mathbb{R},X_e)$ belongs to $X_{1loc}(\mathbb{R})\cap X_{2loc}(\mathbb{R})$. We obtain the uniqueness of the solution to (\ref{771}) with $u, v$ in $\mathcal{C}(\mathbb{R},X_e)$, actually in $L^{\infty}_{loc}(\mathbb{R},X_e)$.\hfill $\Box$

\section{Regular results on (\ref{1}) when $d=3$, $(\alpha,\beta)=(0,2)$ and when $d=4$, $(\alpha,\beta)=(0,0)$}
\qquad In this section, we discuss the regular results on (\ref{1}) when $d=3$, $(\alpha,\beta)=(0,2)$ and when $d=4$, $(\alpha,\beta)=(0,0)$.
First, we give some notations which were used in \cite{Shatah1993, Struwe}. Let $z=(x,t)$ denote the point in space-time $\mathbb{R}^d\times \mathbb{R}$.
\begin{align*}
K(z_0)=\{z=(x,t)||x-x_0|\leq t_0-t\}
\end{align*}
denote the backward light cone with vertex at $z_0=(x_0,t_0)\in \mathbb{R}^d\times \mathbb{R}$,
\begin{align*}
M(z_0)=\{z=(x,t)| |x-x_0|=t_0-t\}
\end{align*}
its mantle, and
\begin{align*}
D(t;z_0)=\{z=(x,t)\in K(z_0)\}
\end{align*}
its spacelike sections for $t$ fixed. For $S<T$ and $Q\subset \mathbb{R}^d\times \mathbb{R}$, let
\begin{align*}
Q^T_S=\{z=(x,t)\in Q|S\leq t\leq T \}
\end{align*}
be the truncated region, while
\begin{align*}
\partial K^T_S=D(S)\cup D(T) \cup M^T_S.
\end{align*}

Given a vector-value function $(u,v)$ on a cone $K(z_0)$, let
\begin{align*}
e_w(u,v)=\frac{1}{2}[\mu(\alpha+2)(|u_t|^2+|\nabla u|^2)+\lambda(\beta+2)(|v_t|^2+|\nabla v|^2)]+\lambda\mu |u|^{\alpha+2}|v|^{\beta+2}
\end{align*}
be its weighted energy density and
\begin{align*}
E_w(u,v,D(t,z_0))=\int_{D(t;z_0)}e_w(u,v)dx
\end{align*}
be its local weighted energy in a ball,
\begin{align*}
d_{z_0}(u,v)=\frac{1}{2}[\mu(\alpha+2)|\frac{y}{|y|}u_t-\nabla u|^2+\lambda(\beta+2)|\frac{y}{|y|}v_t-\nabla v|^2]+\lambda \mu |u|^{\alpha+2}|v|^{\beta+2}
\end{align*}
be the flux density with respect to $M(z_0)$, where $y=x-x_0$, and
\begin{align*}
Flux(u,v, M^T_S(z_0))=\int_{M^T_S(z_0)}d_{z_0}(u,v) do.
\end{align*}

{\bf Lemma 4.1.} {\it Let $(u,v)$ be a regular solution of (\ref{1}) on $K(z_0)$. Then
\begin{align}
E_w(u,v,D(T;z_0))+\frac{1}{\sqrt{2}}Flux(u,v,M^T_S(z_0))=E_w(u,v, D(S;z_0))\label{818xj}
\end{align}
for $0<S\leq T\leq t_0$.}

{\bf Proof:} Multiplying the equation of $u$ by $\mu(\alpha+2)u_t$ and that of $v$ by $\lambda(\beta+2)v_t$ in (\ref{1}), we get
\begin{align}
&\quad\mu(\alpha+2)(u_{tt}-\Delta u+\lambda |u|^{\alpha}|v|^{\beta+2}u)u_t+\lambda(\beta+2)(v_{tt}-\Delta v+\mu|u|^{\alpha+2}|v|^{\beta}v)v_t\nonumber\\
&=\frac{d}{dt}e_w(u,v)-div[\mu(\alpha+2)\nabla u u_t+\lambda(\beta+2) \nabla v v_t]=0.\label{8181}
\end{align}
Let $y=x-x_0$ and $$\eta=\frac{1}{\sqrt{2}}(\frac{y}{|y|},1)$$
be the outward unit normal on $M(z_0)$. Then the weighted energy flux through $M(z_0)$ is given by
\begin{align}
&\qquad\eta\cdot[(-\mu(\alpha+2)\nabla u u_t-\lambda(\beta+2)\nabla v v_t), e_w(u,v)]\nonumber\\
&=\frac{1}{\sqrt{2}}\{\frac{\mu(\alpha+2)}{2}[|\partial_t u|^2-\frac{2y}{|y|}\cdot \nabla u u_t+|\nabla u|^2]+\frac{\lambda(\beta+2)}{2}[|\partial_t v|^2-\frac{2y}{|y|}\cdot \nabla v v_t+|\nabla v|^2]\nonumber\\
&\qquad\qquad+\lambda\mu|u|^{\alpha+2}|v|^{\beta+2}\}\nonumber\\
&=\frac{1}{\sqrt{2}}d_{z_0}(u,v).\label{8182}
\end{align}
Integrating (\ref{8181}) over $K^s_t(z_0)$ and using (\ref{8182}), we obtain (\ref{818xj}).\hfill $\Box$

{\bf Lemma 4.2.} {\it
Let $(u,v)$ be a regular solution of (\ref{1}) on $K(z_0)\setminus \{z_0\}$. Then
\begin{align}
\int_{D(S;z_0)}|u|^{\alpha+2}|v|^{\beta+2}dx\rightarrow 0\quad {\rm as }\quad S\rightarrow t_0.\label{8183}
\end{align}
}

{\bf Proof:} We can shift $z_0$ to the origin and use Morawetz identity to prove the lemma.
Multiplying the equation of $u$ by $\mu(\alpha+2)[tu_t+x\cdot \nabla u+\frac{(d-1)u}{2}]$ and that of $v$ by $\lambda(\beta+2)[tv_t+x\cdot \nabla v+\frac{(d-1)v}{2}]$, summing them up, we get
\begin{align}
\partial_t\left(tQ_0+\frac{d-1}{2}[\mu(\alpha+2)uu_t+\lambda(\beta+2)vv_t]\right)-div(tP_0)+\frac{\lambda\mu}{d}|u|^{\alpha+2}|v|^{\beta+2}=0,\label{818x1}
\end{align}
where
\begin{align*}
Q_0&=e_w(u,v)+[\mu(\alpha+2)(\frac{x}{t}\cdot\nabla u)u_t+\lambda(\beta+2)(\frac{x}{t}\cdot\nabla v)v_t],\nonumber\\
P_0&=\frac{x}{t}\left(\frac{\mu(\alpha+2)(|u_t|^2-|\nabla u|^2)+\lambda(\beta+2)(|v_t|^2-|\nabla v|^2)}{2}-\lambda\mu|u|^{\alpha+2}|v|^{\beta+2}\right)\nonumber\\
&\quad+\mu(\alpha+2)\nabla u\left(u_t+\frac{x}{t}\cdot\nabla u+\frac{(d-1)u}{2t}\right)+\lambda(\beta+2)\nabla v\left(v_t+\frac{x}{t}\cdot\nabla v+\frac{(d-1)v}{2t}\right).
\end{align*}
Integrating (\ref{818x1}) over the truncated cone $K^T_S$ and letting $T\rightarrow 0$, we get
\begin{align}
0&=-\int_{D(S)}\left(SQ_0+\frac{(d-1)}{2}[\mu(\alpha+2)uu_t+\lambda(\beta+2)vv_t]\right)dx+\int_{K_S}\lambda\mu|u|^{\alpha+2}|v|^{\beta+2}dxdt\nonumber\\
&\qquad+\frac{1}{\sqrt{2}}\int_{M_S}\left(tQ_0+x\cdot P_0+\frac{(d-1)}{2}[\mu(\alpha+2)uu_t+\lambda(\beta+2)vv_t]\right)do\nonumber\\
&:=(I)+(II)+(III).\label{818x2}
\end{align}
Obviously,
\begin{align}
(II)\geq 0. \label{920w1}
\end{align}
On $M_S$, we have $|x|=-t$ and
\begin{align*}
(III)&=\frac{1}{\sqrt{2}}\int_{M_S}-|x|\left(\mu(\alpha+2)|u_t-\frac{x\cdot \nabla u}{|x|}|^2+\lambda(\beta+2)|v_t-\frac{x\cdot \nabla v}{|x|}|^2\right)do\nonumber\\
&\quad+\frac{1}{\sqrt{2}}\int_{M_S}\frac{(d-1)}{2}\left(\mu(\alpha+2)u(u_t-\frac{x\cdot \nabla u}{|x|}+\lambda(\beta+2)v(v_t-\frac{x\cdot \nabla v}{|x|}\right)do.
\end{align*}
Now parameterizing $M_S$ by $y\rightarrow (y,-|y|)$ and letting $\tilde{v}(y)=v(y,-|y|)$, we have $do=\sqrt{2}dy$, $y\cdot \frac{\nabla\tilde{u}}{|y|}=\frac{x\cdot \nabla u}{|x|}-u_t$,
and $y\cdot \frac{\nabla\tilde{v}}{|y|}=\frac{x\cdot \nabla v}{|x|}-v_t$. Then
\begin{align*}
(III)&=-\int_{D_S}\frac{\mu(\alpha+2)|y\cdot\nabla \tilde{u}|^2+\lambda(\beta+2)|y\cdot\nabla \tilde{v}|^2}{|y|}dy\nonumber\\
&\quad -\frac{(d-1)}{2}\int_{D_S}[\mu(\alpha+2)\tilde{u}\frac{y\cdot\nabla\tilde{u}}{|y|}+\lambda(\beta+2)\tilde{v}\frac{y\cdot\nabla\tilde{v}}{|y|}] dy\nonumber\\
&=-\int_{D_S}|y|^{-1}\left\{|y\cdot\nabla \tilde{u}+\frac{d-1}{2}\tilde{u}|^2-\frac{(d-1)^2}{4}\tilde{u}^2+|y\cdot\nabla \tilde{v}+\frac{d-1}{2}\tilde{v}|^2-\frac{(d-1)^2}{4}\tilde{v}^2 \right\}dy\nonumber\\
&\quad+\frac{d-1}{4}\int_{D_S}\frac{y\cdot\nabla (\tilde{u})^2+y\cdot \nabla (\tilde{v})^2}{|y|}dy.
\end{align*}
Integrating by parts and using the original coordinates again, we have
\begin{align*}
(III)&=\frac{1}{\sqrt{2}}\int_{M_S}\frac{1}{t}\left(|tu_t+x\cdot\nabla u+\frac{d-1}{2}u|^2+|tv_t+x\cdot\nabla v+\frac{d-1}{2}v|^2\right)do\nonumber\\
&\quad +\frac{d-1}{2}\int_{\partial D(S)}(u^2+v^2)do\nonumber\\
&\geq So(1)+\frac{d-1}{4}\int_{\partial D(S)}(u^2+v^2)do.
\end{align*}
That is,
\begin{align}
-(III)\leq -So(1)-\frac{d-1}{4}\int_{\partial D(S)}(u^2+v^2)do.\label{9201}
\end{align}
Meanwhile,
\begin{align}
(I)&=-\int_{D(S)}S\left[\frac{1}{2}|u_t|^2+\frac{1}{2}\left|\nabla u+\frac{(d-1)}{2}\frac{xu}{|x|^2}\right|^2-\frac{(d-1)}{2}\frac{x\cdot \nabla u}{|x|^2}u-\frac{1}{2}(\frac{d-1}{2})^2\frac{u^2}{|x|^2}\right]dx\nonumber\\
&\quad-\int_{D(S)}S\left[\frac{1}{2}|v_t|^2+\frac{1}{2}\left|\nabla v+\frac{(d-1)}{2}\frac{xv}{|x|^2}\right|^2-\frac{(d-1)}{2}\frac{x\cdot \nabla v}{|x|^2}v-\frac{1}{2}(\frac{d-1}{2})^2\frac{v^2}{|x|^2}\right]dx\nonumber\\
&\quad-\int_{D(S)}\lambda \mu |u|^{\alpha+2}|v|^{\beta+2}dx-\int_{D(S)}[u_t(x\cdot\nabla u+\frac{(d-1)}{2}u)+v_t(x\cdot\nabla v+\frac{(d-1)}{2}v)]dx\nonumber\\
&=-\int_{D(S)}S\left[\frac{1}{2}|u_t|^2+\frac{1}{2}\left|\nabla u+\frac{(d-1)}{2}\frac{xu}{|x|^2}\right|^2+\frac{(d-1)(d-3)}{8}\frac{u^2}{|x|^2}\right]dx\nonumber\\
&\quad-\int_{D(S)}S\left[\frac{1}{2}|v_t|^2+\frac{1}{2}\left|\nabla v+\frac{(d-1)}{2}\frac{xv}{|x|^2}\right|^2+\frac{(d-1)(d-3)}{8}\frac{v^2}{|x|^2}\right]dx\nonumber\\
&\quad-\int_{D(S)}[u_t(x\cdot\nabla u+\frac{(d-1)}{2}u)+v_t(x\cdot\nabla v+\frac{(d-1)}{2}v)]dx-\int_{D(S)}\lambda \mu |u|^{\alpha+2}|v|^{\beta+2}dx\nonumber\\
&\quad-\frac{d-1}{4}\int_{\partial D(S)}(u^2+v^2)do\nonumber\\
&\geq -\int_{D(S)}S\lambda\mu |u|^{\alpha+2}|v|^{\beta+2}dx-\frac{d-1}{4}\int_{\partial D(S)}(u^2+v^2)do.\label{1003w1}
\end{align}
Using (\ref{818x2})--(\ref{1003w1}), we can get that
\begin{align*}
-S\int_{D(S)}\lambda\mu |u|^{\alpha+2}|v|^{\beta+2}dx+So(1)\leq 0,
\end{align*}
which implies that
\begin{align*}
\int_{D(S)} |u|^{\alpha+2}|v|^{\beta+2}dx\rightarrow 0\quad {\rm as}\quad  S\rightarrow t_0. \qquad \qquad\qquad\qquad \qquad\qquad \Box
\end{align*}

Let $\|\cdot\|_{q,s,\tau}$  denote the norm in the space $L^q([s,\tau]; \dot{B}_q^{\frac{1}{2}}(D(t;Z_0)))$, $\dot{B}^{\frac{1}{2}}_q(\mathbb{R}^d)$ be Besov space and
$\dot{B}^{\frac{1}{2}}_q(D(t))$ be the local Besov space, $q<2d$.

{\bf Proposition 4.3} {\it Suppose that $(u,v)$ is a classical solution to equations
(\ref{1}) on $K(z_0)\setminus\{z_0\}$. Then $(u,v)$ is bounded in $L^q([O, t_0]; \dot{B}^{\frac{1}{2}}_q(D(t;z_0)))$ and
\begin{align}
\|u\|_{q,0,t_0}<C(z_0, E(u,D(t_0; z_0))).\label{10041}
\end{align}
}

{\bf Proof:} Similar to (1.8) in \cite{Shatah1993}, for any $p\in [1,\infty]$, we have
\begin{align}
&\quad\|u\|_{L^q([s,\tau];\dot{B}_q^{\frac{1}{2}}(D(t,z_0)))}+ \|v\|_{L^q([s,\tau];\dot{B}_q^{\frac{1}{2}}(D(t,z_0)))}\nonumber\\
&\leq C[E_w(u,v;D(s,z_0))]^{\frac{1}{2}}+\||u|^{\alpha}|v|^{\beta+2}u\|_{L^p([s,\tau];\dot{B}_p^{\frac{1}{2}}(D(t,z_0)))}\nonumber\\
&\quad +\||u|^{\alpha+2}|v|^{\beta}v\|_{L^p([s,\tau];\dot{B}_p^{\frac{1}{2}}(D(t,z_0)))}\label{10051}
\end{align}
and
\begin{align}
\|u\|_{q,s,\tau}+\|v\|_{q,s,\tau}&\lesssim [E_w(u,v;D(s,z_0))]^{\frac{1}{2}}+\||u|^{\alpha+\beta+3}\|_{p,s,\tau}+\||v|^{\alpha+\beta+3}\|_{p,s,\tau},\label{1005x1}\\
\||u|^{\alpha+\beta+3}\|_{p,s,\tau}&\lesssim \||\tilde{u}|^{\alpha+\beta+3}\|_{p,s,\tau}\lesssim \||\tilde{u}|^{\alpha+\beta+2}\|_{L^{p_1}}\|\tilde{u}\|_{q,s,\tau}\nonumber\\
&\lesssim \||u|^{\alpha+\beta+2}\|_{L^{p_1}(K^{\tau}_s(z_0))}\|u\|_{q,s,\tau}\label{1005x2}\\
\||v|^{\alpha+\beta+3}\|_{p,s,\tau}&\lesssim \||\tilde{v}|^{\alpha+\beta+3}\|_{p,s,\tau}\lesssim \||\tilde{v}|^{\alpha+\beta+2}\|_{L^{p_1}}\|\tilde{v}\|_{q,s,\tau}\nonumber\\
&\lesssim \||v|^{\alpha+\beta+2}\|_{L^{p_1}(K^{\tau}_s(z_0))}\|v\|_{q,s,\tau}\label{1005x3}\\
\||u|^{\alpha+\beta+2}\|_{L^{p_1}}&=\|u\|^{\alpha+\beta+2}_{L^{p_2}}, \||v|^{\alpha+\beta+2}\|_{L^{p_1}}=\|v\|^{\alpha+\beta+2}_{L^{p_2}}.\label{1005x4}
\end{align}
Here
\begin{align}
p_1=\frac{(d+1)}{2},\quad p_2=\frac{2(d+1)}{d-2}.
\end{align}
By Sobolev embedding,
\begin{align*}
\|u\|_{L^{p_2}(D(t;z_0))}\lesssim \|u\|^{\theta}_{\dot{B}^{\frac{1}{2}}_q(D(t;z_0))}\|u\|^{1-\theta}_{L^{\alpha+\beta+4}(D(t;z_0))}, \quad \theta=\frac{d-2}{d-1}\\
\|v\|_{L^{p_2}(D(t;z_0))}\lesssim \|v\|^{\theta}_{\dot{B}^{\frac{1}{2}}_q(D(t;z_0))}\|v\|^{1-\theta}_{L^{\alpha+\beta+4}(D(t;z_0))},
\end{align*}
then (\ref{1005x1}) implies that
\begin{align}
\|u\|_{q,s,\tau}+\|v\|_{q,s,\tau}&\lesssim [E_w(u,v;D(s,z_0))]^{\frac{1}{2}}+\sup_{s\leq t\leq \tau}\|u\|^{\vartheta}_{L^{\alpha+\beta+4}(D(t;z_0))}\|u\|_{q,s,\tau}^{\gamma}\nonumber\\
&\quad+
\sup_{s\leq t\leq \tau}\|v\|^{\vartheta}_{L^{\alpha+\beta+4}(D(t;z_0))}\|v\|_{q,s,\tau}^{\gamma}.
\end{align}
By Lemma 4.2, we have
\begin{align}
\sup_{s\leq t\leq \tau}\|u\|^{\vartheta}_{L^{\alpha+\beta+4}(D(t;z_0))}+
\sup_{s\leq t\leq \tau}\|u\|^{\vartheta}_{L^{\alpha+\beta+4}(D(t;z_0))}\leq \epsilon.\label{1005x5}
\end{align}
Note that $E_w(u,v; D(s;z_0))$ is bounded by the
initial energy $E_w(u_0,v_0)$. This implies that for any $\epsilon>0$, and  $s$ close to $t_0$ and
$s<r<t_0$, we have
\begin{align}
\|u\|_{q,s,\tau}+\|v\|_{q,s,\tau}\leq C(E_w(u_0,v_0))+\epsilon\|u\|^{\gamma}_{q,s,\tau}+\|v\|^{\gamma}_{q,s,\tau}.
\end{align}
Choosing $\epsilon<\epsilon_02^{-\gamma}[C(E_w(u_0,v_0))]^{1-\gamma}$, we can obtain (\ref{10041}).\hfill $\Box$

{\bf Theorem 4.4.} {\it Suppose that $(u,v)$ is a solution to equations (\ref{1}) with
smooth initial data. If $d=3$,  $(\alpha,\beta)=(0,2)$ or $d=4$, $(\alpha,\beta)=(0,0)$, then $(u,v)$ is regular.}

{\bf Proof:} Without loss of generality, we suppose that $(u_{10},u_{20})$ and $(v_{10},v_{20})$ have compact supports by finite propagation speed. Let  $(u,v)\in C^{\infty}(R^d\times [0, t_0])$ be the unique maximal
solution of (\ref{1}) and consider $x_0\in R^d$. Then $(u,v)$ may be extended
smoothly to a neighborhood of $z_0=(x_0,t_0)$. By differentiating equations
(\ref{1}) and using Strichartz estimate
\begin{align}
\|u\|_{L^q(R^{d+1})}&\lesssim \|u_{10}\|_{H^{\frac{1}{2},2}(\mathbb{R}^d)}+\|u_{20}\|_{H^{-\frac{1}{2},2}(\mathbb{R}^d)}+\||u|^{\alpha}|v|^{\beta+2}u\|_{L^p(R^{d+1})},\nonumber\\
\|v\|_{L^q(R^{d+1})}&\lesssim \|v_{10}\|_{H^{\frac{1}{2},2}(\mathbb{R}^d)}+\|v_{20}\|_{H^{-\frac{1}{2},2}(\mathbb{R}^d)}+\||u|^{\alpha+2}|v|^{\beta}v\|_{L^p(R^{d+1})}
\end{align}
we obtain the following estimate for any first-order
derivative:
\begin{align}
&\quad \|Du\|_{L^q(K^{\tau}_s(z_0))}+\|Dv\|_{L^q(K^{\tau}_s(z_0))}\nonumber\\
&\lesssim C(E_w(Du,Dv,D(S;z_0)))+\||u|^{\alpha}|v|^{\beta+2}|Du|\|_{L^p(K^{\tau}_s(z_0))}+\||u|^{\alpha+1}|v|^{\beta+1}|Dv|\|_{L^p(K^{\tau}_s(z_0))}\nonumber\\
&\quad+\||u|^{\alpha+2}|v|^{\beta}|Dv|\|_{L^p(K^{\tau}_s(z_0))}+\||u|^{\alpha+1}|v|^{\beta+1}|Du|\|_{L^p(K^{\tau}_s(z_0))}\nonumber\\
&:=C(E_w(Du,Dv,D(S;z_0)))+(I)+(II)+(III)+(IV).\label{12221}
\end{align}
Similar to the computations in the proof of Proposition 4.3, we can get
\begin{align}
(I)+(II)+(III)+(IV)&\lesssim [\|Du\|_{L^q}+\|Dv\|_{L^q}][\|u\|^{\alpha+\beta+2}_{L^{p_2}(K^{\tau}_s(z_0))}+\|v\|^{\alpha+\beta+2}_{L^{p_2}(K^{\tau}_s(z_0))}]\nonumber\\
&\quad \lesssim \sup_{s\leq t\leq \tau}[\|u\|^{\beta}_{L^{\alpha+\beta+4}(D(t;z_0))}+\|v\|^{\beta}_{L^{\alpha+\beta+4}(D(t;z_0))}][\|u\|^{\gamma-1}_{q,s,\tau}+\|v\|^{\gamma-1}_{q,s,\tau}]\nonumber\\
&\qquad \times [\|Du\|_{L^q(K^{\tau}_s(z_0))}+\|Dv\|_{L^q(K^{\tau}_s(z_0))}].
\end{align}
By Lemma 4.2, the coefficients in front of $[\|Du\|_{L^q(K^{\tau}_s(z_0))}+\|Dv\|_{L^q(K^{\tau}_s(z_0))}]$ on the right hand side of (\ref{12221}) can be small enough, which implies that $Du\in L^q(K(z_0))$ and $Dv\in L^q(K(z_0))$.

Differentiating the equation again, we can get the equations of $D^2u$ and $D^2v$. If $d=3$, $(\alpha,\beta=(0,2)$, there will appear the additional terms $|v|^3DuDv$ and $
|v|^2u|Dv|^2$ in the equation of $D^2u$,  $|u|^3DuDv$ and $|u|^2v|Du|^2$ in the equation of $D^2v$; If $d=4$, $(\alpha,\beta=(0,0)$, there will appear the additional terms $vDuDv$ and $u|Dv|^2$ in the equation of $D^2u$,  $uDuDv$ and $v|Du|^2$ in the equation of $D^2v$. Using H\"{o}lder's inequality and the Sobolev embedding theorem, we obtain
\begin{align}
&\quad\||v|^3DuDv\|_{L^p}+\||v|^2u|Dv|^2\|_{L^p}+\||u|^3DuDv\|_{L^p}+\||u|^2v|Du|^2\|_{L^p}\nonumber\\
&\lesssim [\|Du\|_{L^q}+\|u\|_{L^q}+\|Dv\|_{L^q}+\|v\|_{L^q}]^{\rho}[\|D^2u\|_{L^q}+\|D^2v\|_{L^q}]^{\sigma}
\end{align}
 if $d=3$, $(\alpha,\beta)=(0,2)$ and
\begin{align}
&\quad\|vDuDv\|_{L^p}+\|u|Dv|^2\|_{L^p}+\|uDuDv\|_{L^p}+\|v|Du|^2\|_{L^p}\nonumber\\
&\lesssim [\|Du\|_{L^q}+\|u\|_{L^q}+\|Dv\|_{L^q}+\|v\|_{L^q}]^{\rho}[\|D^2u\|_{L^q}+\|D^2v\|_{L^q}]^{\sigma}
\end{align}
if $d=4$, $(\alpha,\beta)=(0,0)$, where $\sigma\leq 1$. If $d=3$, then  $u, v\in W^{2,q}(K(z_0))$ implies that $|v|^4u, |u|^4v\in H^{2,2}(K(z_0))$, which in turn implies that
$D^3u, D^3v\in L^{\infty}([0,t_0];L^2(D(t;z_0)))$.
Using energy estimate and Grownwall inequality, we can get
\begin{align}
D^ku, D^kv\in L^2([0,t_0];L^2(D(t;z_0))) \quad {\rm for \ all} \ k
\end{align}
and
\begin{align}
E_w(u,v,D(t;z_0))\rightarrow 0\quad {\rm as}\quad t\rightarrow t_0.
\end{align}

Consequently, for given $\epsilon>0$, $E_w(u,v,D(s;z_0))<\epsilon$ for some $s<t_0$. Since $(u,v)$ is smooth in the neighbourhood of $D(s;z_0)$, hence $E_w(u,v,D(s;\bar{z}))<\epsilon$,
\begin{align}
\sup_{s\leq t}[\|u\|_{L^6(D(s;\bar{z}))}+\|v\|_{L^6(D(s;\bar{z}))}]<\epsilon \quad {\rm when}\ d=3
\end{align}
and
\begin{align}
\sup_{s\leq t}[\|u\|_{L^4(D(s;\bar{z}))}+\|v\|_{L^4(D(s;\bar{z}))}]<\epsilon \quad {\rm when}\ d=4
\end{align}
keep true for $\bar{z}$ in a neighborhood of $z_0$. By the standard argument similar to Remark 3.4 in \cite{Struwe}, $(u,v)$ can be smoothly extended to
this neighborhood.

\section{$\dot{H}^1\times L^2$ scattering results when $(\alpha,\beta)=(0,2)$ in dimension 3 and $(\alpha,\beta)=(0,0)$ in dimension 4}
\qquad In this section, we focus on the wellposedness and scattering for the solution $(u,v)$ of (\ref{771}) with $(u,u_t), (v,v_t)\in \dot{H}^1\times L^2$
when $\alpha+\beta=2$ in dimension 3 and $(\alpha,\beta)=(0,0)$ in dimension 4.

First, we prove the following local wellposedness result.

{\bf Theorem 5.1.} {\it Let $(\alpha,\beta)=(0,2)$, $\rho=5$ when $d=3$ and $(\alpha,\beta)=(0,0)$, $\rho=3$ when $d=4$, $(u_{10},u_{20}), (v_{10},v_{20})\in \dot{H}^1(\mathbb{R}^d)\times L^2(\mathbb{R}^d)$ satisfying
\begin{align}
\|u_{10}\|_2+\|\nabla u_{10}\|_2+\|u_{20}\|_2+\|v_{10}\|_2+\|\nabla v_{10}\|_2+\|v_{20}\|_2<\delta,\label{8071}
\end{align}
where $\delta>0$ is sufficiently small. Then there exists a unique solution $(u,v)$ of (\ref{771}) and satisfies
$(u,u_t), (v, v_t)\in V\oplus L^{\infty}(\mathbb{R},L^2(\mathbb{R}^d))$, where $V:=L^{\rho}(\mathbb{R}, \dot{H}^{s,r}(\mathbb{R}^d))\cap L^{\infty}(\mathbb{R}, \dot{H}^1(\mathbb{R}^d))$ with
\begin{align}
s:=\frac{(d-1)\rho-(d+1)}{(d-1)\rho},\quad r:=\frac{2(d-1)\rho}{(d-1)\rho-4}.
\end{align}
}

{\bf Proof:} By the results of Theorem 2 in \cite{Pecher1984} and its Corollary, we can choose $q=\frac{2(d-1)\rho}{(d-1)\rho-4}$(because $2<\rho<\infty$, $d=3$ or $d=4$) such that
\begin{align*}
&\dot{K}(t)u_{10}+K(t)u_{20}\in V,\quad  \dot{K}(t)v_{10}+K(t)v_{20}\in V,\\
&\|\dot{K}(t)u_{10}+K(t)u_{20}; V\|\leq c\delta,\quad  \|\dot{K}(t)v_{10}+K(t)v_{20}; V\|\leq c\delta.
\end{align*}
For $(u_1,v_1), (u_2,v_2)\in V\oplus V$, using the decay estimate as that of Theorem 0(a) in \cite{Pecher}, we have
\begin{align}
&\quad\|\int^t_{-\infty}\frac{\sin[\omega(t-\tau)]}{\omega}[\lambda|u_1|^{\alpha}|v_1|^{\beta+2}u_1(\tau)-\lambda|u_2|^{\alpha}|v_2|^{\beta+2}u_2(\tau)]
d\tau\|_{\dot{H}^{s,r}}\nonumber\\
&\quad+\|\int^t_{-\infty}\frac{\sin[\omega(t-\tau)]}{\omega}[\mu|u_1|^{\alpha+2}|v_1|^{\beta}v_1(\tau)-\mu|u_2|^{\alpha+2}|v_2|^{\beta}v_2(\tau)]
d\tau\|_{\dot{H}^{s,r}}\nonumber\\
&\leq c\int^t_{-\infty}(t-\tau)^{-\frac{2}{\rho}}\|[\lambda|u_1|^{\alpha}|v_1|^{\beta+2}u_1(\tau)-\lambda|u_2|^{\alpha}|v_2|^{\beta+2}u_2(\tau)]
\|_{\dot{H}^{\bar{s},r'}}d\tau \nonumber\\
&\quad+ c\int^t_{-\infty}(t-\tau)^{-\frac{2}{\rho}}\|[\mu|u_1|^{\alpha+2}|v_1|^{\beta}v_1(\tau)-\mu|u_2|^{\alpha+2}|v_2|^{\beta}v_2(\tau)]
\|_{\dot{H}^{\bar{s},r'}}d\tau, \label{809w1}
\end{align}
where $\bar{s}=\frac{d+1}{(d-1)\rho}$. Letting $\frac{1}{\tilde{r}'}=\frac{(d+2)\rho+2}{2d\rho}$, we know that the imbedding $\dot{H}^{1,\tilde{r}'}\subset \dot{H}^{\bar{s},r'}$ holds. We will use Theorem 1 on Page 119 in \cite{Stein} to establish the following estitimates.

If $d=3$, then
\begin{align}
&\quad \lambda\|[|u_1|^{\alpha}|v_1|^{\beta+2}u_1(\tau)-|u_2|^{\alpha}|v_2|^{\beta+2}u_2(\tau)]
\|_{\dot{H}^{\bar{s},r'}}+ \mu\|[|u_1|^{\alpha+2}|v_1|^{\beta}v_1(\tau)-|u_2|^{\alpha+2}|v_2|^{\beta}v_2(\tau)]
\|_{\dot{H}^{\bar{s},r'}}\nonumber\\
&\leq c\sum_{|k|=1}\||v_1|^4D^ku_1+u_1|v_1|^3D^kv_1-|v_2|^4D^ku_2-u_2|v_2|^3D^kv_2\|_{L^{\tilde{r}'}}\nonumber\\
&\quad +c\sum_{|k|=1}\||u_1|^4D^kv_1+v_1|u_1|^3D^ku_1-|u_2|^4D^kv_2-v_2|u_2|^3D^ku_2\|_{L^{\tilde{r}'}}\nonumber\\
&\leq c\sum_{|k|=1}\||v_1-v_2|(|v_1|^3+|v_2|^3)D^ku_1\|_{L^{\tilde{r}'}}+\||u_1-u_2||v_1|^3D^kv_1\|_{L^{\tilde{r}'}}\nonumber\\
&\quad+\||v_1-v_2||u_2|(|v_1|^2+|v_2|^2)D^kv_2\|_{L^{\tilde{r}'}}+\||v_2|^4(D^ku_1-D^ku_2)\|_{L^{\tilde{r}'}}+\||u_2||v_2|^3(D^kv_1-D^kv_2)\|_{L^{\tilde{r}'}}\nonumber\\
&\quad +c\sum_{|k|=1}\||u_1-u_2|(|u_1|^3+|u_2|^3)D^kv_1\|_{L^{\tilde{r}'}}+\||v_1-v_2||u_1|^3D^ku_1\|_{L^{\tilde{r}'}}\nonumber\\
&\quad+\||u_1-u_2||v_2|(|u_1|^2+|u_2|^2)D^ku_2\|_{L^{\tilde{r}'}}+\||u_2|^4(D^kv_1-D^kv_2)\|_{L^{\tilde{r}'}}+\||v_2||u_2|^3(D^ku_1-D^ku_2)\|_{L^{\tilde{r}'}}\nonumber\\
&:=(1)+(2)+(3)+(4)+(5)+(6)+(7)+(8)+(9)+(10).\label{8091}
\end{align}

If $d=4$, then
\begin{align}
&\quad \lambda\|[|u_1|^{\alpha}|v_1|^{\beta+2}u_1(\tau)-|u_2|^{\alpha}|v_2|^{\beta+2}u_2(\tau)]
\|_{\dot{H}^{\bar{s},r'}}+ \mu\|[|u_1|^{\alpha+2}|v_1|^{\beta}v_1(\tau)-|u_2|^{\alpha+2}|v_2|^{\beta}v_2(\tau)]
\|_{\dot{H}^{\bar{s},r'}}\nonumber\\
&\leq c\sum_{|k|=1}\||v_1|^2D^ku_1+u_1v_1D^kv_1-|v_2|^2D^ku_2-u_2v_2D^kv_2\|_{L^{\tilde{r}'}}\nonumber\\
&\quad +c\sum_{|k|=1}\||u_1|^2D^kv_1+v_1u_1D^ku_1-|u_2|^2D^kv_2-v_2u_2D^ku_2\|_{L^{\tilde{r}'}}\nonumber\\
&\leq c\sum_{|k|=1}\||v_1-v_2|(|v_1|+|v_2|)D^ku_1\|_{L^{\tilde{r}'}}+\||u_1-u_2||v_1|D^kv_1\|_{L^{\tilde{r}'}}\nonumber\\
&\quad+\||v_1-v_2||u_2|D^kv_2\|_{L^{\tilde{r}'}}+\||v_2|^2(D^ku_1-D^ku_2)\|_{L^{\tilde{r}'}}+\||u_2||v_2|(D^kv_1-D^kv_2)\|_{L^{\tilde{r}'}}\nonumber\\
&\quad +c\sum_{|k|=1}\||u_1-u_2|(|u_1|+|u_2|)D^kv_1\|_{L^{\tilde{r}'}}+\||v_1-v_2||u_1|D^ku_1\|_{L^{\tilde{r}'}}\nonumber\\
&\quad+\||u_1-u_2||v_2|D^ku_2\|_{L^{\tilde{r}'}}+\||u_2|^2(D^kv_1-D^kv_2)\|_{L^{\tilde{r}'}}+\||v_2||u_2|(D^ku_1-D^ku_2)\|_{L^{\tilde{r}'}}\nonumber\\
&:=(1)'+(2)'+(3)'+(4)'+(5)'+(6)'+(7)'+(8)'+(9)'+(10)'.\label{8091'}
\end{align}
Using H\"{o}lder's inequality, (1), (2), (3), (6),(7) and (8) in (\ref{8091}), as well as ${\rm (1)', (2)', (3)', (6)', (7)'}$ and ${\rm (8)'}$ in (\ref{8091'}) can be estimated by \begin{align}
&[\|u_1-u_2\|_{L^{\tilde{r}'\hat{p}}}+\|v_1-v_2\|_{L^{\tilde{r}'\hat{p}}}][\|u_1\|^{\rho-2}_{L^{\tilde{r}'(\rho-2)\hat{q}}}+\|u_2\|^{\rho-2}_{L^{\tilde{r}'(\rho-2)\hat{q}}}
+\|v_1\|^{\rho-2}_{L^{\tilde{r}'(\rho-2)\hat{q}}}+\|v_2\|^{\rho-2}_{L^{\tilde{r}'(\rho-2)\hat{q}}}]\nonumber\\
&\times [\|D^ku_1\|_{L^{\tilde{r}'\hat{r}}}+\|D^kv_1\|_{L^{\tilde{r}'\hat{r}}}+\|D^ku_2\|_{L^{\tilde{r}'\hat{r}}}+\|D^kv_2\|_{L^{\tilde{r}'\hat{r}}}].\label{8092}
\end{align}
Since the choice $\hat{r}=\frac{2}{\tilde{r}'}$ and $\hat{p}=(\rho-2)\hat{q}$ can lead to $\tilde{r}'\hat{p}=\tilde{r}'(\rho-2)\hat{q}=\frac{d\rho(\rho-1)}{\rho+1}$, and the embedding $\dot{H}^{s,r}\subset L^{\frac{d\rho(\rho-1)}{\rho+1}}$ holds if and only if $\rho=\frac{d+2}{d-2}$, we get
\begin{align}
&\quad \lambda\|[|u_1|^{\alpha}|v_1|^{\beta+2}u_1(\tau)-|u_2|^{\alpha}|v_2|^{\beta+2}u_2(\tau)]
\|_{\dot{H}^{\bar{s},r'}}+ \mu\|[|u_1|^{\alpha+2}|v_1|^{\beta}v_1(\tau)-|u_2|^{\alpha+2}|v_2|^{\beta}v_2(\tau)]
\|_{\dot{H}^{\bar{s},r'}}\nonumber\\
&\leq c[\|u_1-u_2\|_{\dot{H}^{s,r}}+\|v_1-v_2\|_{\dot{H}^{s,r}}][\|u_1\|^{\rho-2}_{\dot{H}^{s,r}}+\|u_2\|^{\rho-2}_{\dot{H}^{s,r}}
+\|v_1\|^{\rho-2}_{\dot{H}^{s,r}}+\|v_2\|^{\rho-2}_{\dot{H}^{s,r}}]\nonumber\\
&\quad \times [\|u_1\|_{\dot{H}^{1,2}}+\|u_2\|_{\dot{H}^{1,2}}+\|v_1\|_{\dot{H}^{1,2}}+\|v_2\|_{\dot{H}^{1,2}}]\nonumber\\
&\quad +c[\|u_1\|^{\rho-1}_{\dot{H}^{s,r}}+\|u_2\|^{\rho-1}_{\dot{H}^{s,r}}+\|v_1\|^{\rho-1}_{\dot{H}^{s,r}}+\|v_2\|^{\rho-1}_{\dot{H}^{s,r}}]
[\|u_1-u_2\|_{\dot{H}^{1,2}}+\|v_1-v_2\|_{\dot{H}^{1,2}}].\label{8093}
\end{align}
Then we can obtain the estimate for the left hand side of (\ref{809w1}) as follows:
\begin{align}
&\quad LHS(\ref{809w1})\nonumber\\
&\leq c\int^t_{-\infty}(t-\tau)^{-\frac{2}{\rho}}[\|u_1-u_2\|_{\dot{H}^{s,r}}+\|v_1-v_2\|_{\dot{H}^{s,r}}][\|u_1\|^{\rho-2}_{\dot{H}^{s,r}}+\|u_2\|^{\rho-2}_{\dot{H}^{s,r}}
+\|v_1\|^{\rho-2}_{\dot{H}^{s,r}}+\|v_2\|^{\rho-2}_{\dot{H}^{s,r}}]\nonumber\\
&\quad \times [\|u_1\|_{L^{\infty}(\mathbb{R},\dot{H}^{1,2}(\mathbb{R}^d))}+\|u_2\|_{L^{\infty}(\mathbb{R},\dot{H}^{1,2}(\mathbb{R}^d))}
+\|v_1\|_{L^{\infty}(\mathbb{R},\dot{H}^{1,2}(\mathbb{R}^d))}+\|v_2\|_{L^{\infty}(\mathbb{R},\dot{H}^{1,2}(\mathbb{R}^d))}]d\tau\nonumber\\
&+c\int^t_{-\infty}(t-\tau)^{-\frac{2}{\rho}}[\|u_1\|^{\rho-1}_{\dot{H}^{s,r}}+\|u_2\|^{\rho-1}_{\dot{H}^{s,r}}+\|v_1\|^{\rho-1}_{\dot{H}^{s,r}}
+\|v_2\|^{\rho-1}_{\dot{H}^{s,r}}]\nonumber\\
&\quad \times [\|u_1-u_2\|_{L^{\infty}(\mathbb{R},\dot{H}^{1,2}(\mathbb{R}^d))}+\|v_1-v_2\|_{L^{\infty}(\mathbb{R},\dot{H}^{1,2}(\mathbb{R}^d))}]d\tau.\label{809w3}
\end{align}
Note that for $0<\theta<1$, $\tilde{\rho}=\frac{1}{\theta}$ and $\frac{1}{p}=\frac{1}{q}+\frac{1}{\tilde{\rho}'}$, the following general Young's inequality holds
\begin{align*}
\|\frac{1}{|t|^{\theta}}*g\|_{L^q(\mathbb{R})}\leq c\|g\|_{L^p(\mathbb{R})}.
\end{align*}
Taking $\tilde{\rho}=\frac{\rho}{2}$, $q=\rho$ and $p=\frac{\rho}{\rho-1}$, we have
\begin{align}
&\quad\|\int^t_{-\infty}\frac{\sin[\omega(t-\tau)]}{\omega}[\lambda|u_1|^{\alpha}|v_1|^{\beta+2}u_1(\tau)-\lambda|u_2|^{\alpha}|v_2|^{\beta+2}u_2(\tau)]
d\tau\|_{L^{\rho}(\mathbb{R},\dot{H}^{s,r}(\mathbb{R}^d))}\nonumber\\
&\quad+\|\int^t_{-\infty}\frac{\sin[\omega(t-\tau)]}{\omega}[\mu|u_1|^{\alpha+2}|v_1|^{\beta}v_1(\tau)-\mu|u_2|^{\alpha+2}|v_2|^{\beta}v_2(\tau)]
d\tau\|_{L^{\rho}(\mathbb{R},\dot{H}^{s,r}(\mathbb{R}^d))}\nonumber\\
&\leq c[\|u_1\|_{L^{\infty}(\mathbb{R},\dot{H}^{1,2}(\mathbb{R}^d))}+\|u_2\|_{L^{\infty}(\mathbb{R},\dot{H}^{1,2}(\mathbb{R}^d))}
+\|v_1\|_{L^{\infty}(\mathbb{R},\dot{H}^{1,2}(\mathbb{R}^d))}+\|v_2\|_{L^{\infty}(\mathbb{R},\dot{H}^{1,2}(\mathbb{R}^d))}]\nonumber\\
&\left(\int^{+\infty}_{-\infty}[\|u_1-u_2\|^{\frac{\rho}{\rho-1}}_{\dot{H}^{s,r}}+\|v_1-v_2\|^{\frac{\rho}{\rho-1}}_{\dot{H}^{s,r}}]
[\|u_1\|^{\frac{\rho(\rho-2)}{\rho-1}}_{\dot{H}^{s,r}}+\|u_2\|^{\frac{\rho(\rho-2)}{\rho-1}}_{\dot{H}^{s,r}}+\|v_1\|^{\frac{\rho(\rho-2)}{\rho-1}}_{\dot{H}^{s,r}}
+\|v_2\|^{\frac{\rho(\rho-2)}{\rho-1}}_{\dot{H}^{s,r}}]d\tau\right)^{\frac{\rho-1}{\rho}}\nonumber\\
&\quad +c[\|u_1-u_2\|_{L^{\infty}(\mathbb{R},\dot{H}^{1,2}(\mathbb{R}^d))}+\|v_1-v_2\|_{L^{\infty}(\mathbb{R},\dot{H}^{1,2}(\mathbb{R}^d))}]\nonumber\\
&\qquad \times\left(\int^{+\infty}_{-\infty}[\|u_1\|^{\rho}_{\dot{H}^{s,r}}+\|u_2\|^{\rho}_{\dot{H}^{s,r}}+\|v_1\|^{\rho}_{\dot{H}^{s,r}}
+\|v_2\|^{\rho}_{\dot{H}^{s,r}}]d\tau\right)^{\frac{\rho-1}{\rho}}\nonumber\\
&\leq c[\|u_1-u_2\|_V+\|v_1-v_2\|_V][\|u_1\|^{\rho-1}_V+\|u_2\|^{\rho-1}_V+\|v_1\|^{\rho-1}_V+\|v_2\|^{\rho-1}_V].\label{8101}
\end{align}
Since the imbedding $\dot{H}^{s,r}\subset L^{2\rho}$ holds if and only if $\rho=\frac{d+2}{d-2}$
\begin{align}
&\quad\|\int^t_{-\infty}\frac{\sin[\omega(t-\tau)]}{\omega}[\lambda|u_1|^{\alpha}|v_1|^{\beta+2}u_1(\tau)-\lambda|u_2|^{\alpha}|v_2|^{\beta+2}u_2(\tau)]
d\tau\|_{L^{\infty}(\mathbb{R},\dot{H}^{1,2}(\mathbb{R}^d))}\nonumber\\
&\quad+\|\int^t_{-\infty}\frac{\sin[\omega(t-\tau)]}{\omega}[\mu|u_1|^{\alpha+2}|v_1|^{\beta}v_1(\tau)-\mu|u_2|^{\alpha+2}|v_2|^{\beta}v_2(\tau)]
d\tau\|_{L^{\infty}(\mathbb{R},\dot{H}^{1,2}(\mathbb{R}^d))}\nonumber\\
&\leq c\int^{+\infty}_{-\infty}\|[|u_1|^{\alpha}|v_1|^{\beta+2}u_1(\tau)-|u_2|^{\alpha}|v_2|^{\beta+2}u_2(\tau)]\|_{L^2(\mathbb{R}^d)}d\tau\nonumber\\
&\quad +c\int^{+\infty}_{-\infty}\|[|u_1|^{\alpha+2}|v_1|^{\beta}v_1(\tau)-|u_2|^{\alpha+2}|v_2|^{\beta}v_2(\tau)]\|_{L^2(\mathbb{R}^d)}d\tau\nonumber\\
&\leq c\int^{+\infty}_{-\infty}[\|u_1\|^{\rho-1}_{L^{2\rho}}+\|u_2\|^{\rho-1}_{L^{2\rho}}+\|v_1\|^{\rho-1}_{L^{2\rho}}+\|v_2\|^{\rho-1}_{L^{2\rho}}]
[\|u_1-u_2\|_{L^{2\rho}}+\|v_1-v_2\|_{L^{2\rho}}]d\tau\nonumber\\
&\leq c[\|u_1\|^{\rho-1}_V+\|u_2\|^{\rho-1}_V+\|v_1\|^{\rho-1}_V+\|v_2\|^{\rho-1}_V][\|u_1-u_2\|_V+\|v_1-v_2\|_V].\label{8102}
\end{align}
Denoting the transformations which map $(u,v)$ into the right hand sides of the equations in (\ref{771}) respectively by $T_1$ and $T_2$, we have shown
\begin{align*}
&\quad \|T_1(u_1,v_1)-T_1(u_2,v_2)\|_V+\|T_2(u_1,v_1)-T_2(u_2,v_2)\|_V\nonumber\\
&\leq c[\|u_1\|^{\rho-1}_V+\|u_2\|^{\rho-1}_V+\|v_1\|^{\rho-1}_V+\|v_2\|^{\rho-1}_V][\|u_1-u_2\|_V+\|v_1-v_2\|_V],
\end{align*}
and
$$
\|T_1(u,v)\|_V+\|T_2(u,v)\|_V\leq c\delta+c[\|u\|^{\rho}_V+\|v\|^{\rho}_V].
$$
If $c\delta\leq \frac{1}{4}\delta_1$ and $4c\delta_1^{\rho-1}\leq \frac{1}{2}$, then $\|u_1\|_V, \|u_2\|_V, \|v_1\|_V, \|v_2\|_V\leq \delta_1$, and
\begin{align*}
&\|T_1(u_1,v_1)-T_1(u_2,v_2)\|_V+\|T_2(u_1,v_1)-T_2(u_2,v_2)\|_V\leq \frac{1}{2}[\|u_1-u_2\|_V+\|v_1-v_2\|_V]\\
&\|T_1(u,v)\|_V+\|T_2(u,v)\|_V\leq \delta_1.
\end{align*}
By the contraction mapping principle, there exists a unique solution $(u,v)$ within the ball $\|u\|_V+\|v\|_V\leq \delta_1$. To show the uniqueness within the whole of $V$, we give the following estimates
\begin{align*}
&\quad \|u_1-u_2\|_{L^{\rho}(I,\dot{H}^{s,r}(\mathbb{R}^d))}+\|v_1-v_2\|_{L^{\rho}(I,\dot{H}^{s,r}(\mathbb{R}^d))}\nonumber\\
&\leq c[\|u_1\|_{L^{\infty}(I,\dot{H}^{1,2}(\mathbb{R}^d))}+\|u_2\|_{L^{\infty}(I,\dot{H}^{1,2}(\mathbb{R}^d))}+\|v_1\|_{L^{\infty}(I,\dot{H}^{1,2}(\mathbb{R}^d))}
+\|v_2\|_{L^{\infty}(I,\dot{H}^{1,2}(\mathbb{R}^d))}]\nonumber\\
&\quad\times[\|u_1\|^{\rho-2}_{L^{\rho}(I,\dot{H}^{s,r}(\mathbb{R}^d))}+\|u_2\|^{\rho-2}_{L^{\rho}(I,\dot{H}^{s,r}(\mathbb{R}^d))}
+\|v_1\|^{\rho-2}_{L^{\rho}(I,\dot{H}^{s,r}(\mathbb{R}^d))}
+\|v_2\|^{\rho-2}_{L^{\rho}(I,\dot{H}^{s,r}(\mathbb{R}^d))}]\nonumber\\
&\quad \times [\|u_1-u_2\|_{L^{\rho}(I,\dot{H}^{s,r}(\mathbb{R}^d))}+\|v_1-v_2\|_{L^{\rho}(I,\dot{H}^{s,r}(\mathbb{R}^d))}]\nonumber\\
&\quad +[\|u_1\|^{\rho-1}_{L^{\rho}(I,\dot{H}^{s,r}(\mathbb{R}^d))}+\|u_2\|^{\rho-1}_{L^{\rho}(I,\dot{H}^{s,r}(\mathbb{R}^d))}
+\|v_1\|^{\rho-1}_{L^{\rho}(I,\dot{H}^{s,r}(\mathbb{R}^d))}
+\|v_2\|^{\rho-1}_{L^{\rho}(I,\dot{H}^{s,r}(\mathbb{R}^d))}]\nonumber\\
&\quad \times [\|u_1-u_2\|_{L^{\infty}(I,\dot{H}^{1,2}(\mathbb{R}^d))}+\|v_1-v_2\|_{L^{\infty}(I,\dot{H}^{1,2}(\mathbb{R}^d))}]\nonumber\\
\end{align*}
and
\begin{align*}
&\quad \|u_1-u_2\|_{L^{\infty}(I,\dot{H}^{s,r}(\mathbb{R}^d))}+\|v_1-v_2\|_{L^{\infty}(I,\dot{H}^{s,r}(\mathbb{R}^d))}\nonumber\\
&\leq [\|u_1\|^{\rho-1}_{L^{\rho}(I,\dot{H}^{s,r}(\mathbb{R}^d))}+\|u_2\|^{\rho-1}_{L^{\rho}(I,\dot{H}^{s,r}(\mathbb{R}^d))}
+\|v_1\|^{\rho-1}_{L^{\rho}(I,\dot{H}^{s,r}(\mathbb{R}^d))}
+\|v_2\|^{\rho-1}_{L^{\rho}(I,\dot{H}^{s,r}(\mathbb{R}^d))}]\nonumber\\
&\quad \times [\|u_1-u_2\|_{L^{\rho}(I,\dot{H}^{s,r}(\mathbb{R}^d))}+\|v_1-v_2\|_{L^{\rho}(I,\dot{H}^{s,r}(\mathbb{R}^d))}],\nonumber\\
\end{align*}
where $I=(-\infty, \bar{T})$.

Since $u_1, u_2, v_1, v_2\in L^{\rho}(\mathbb{R},\dot{H}^{s,r}(\mathbb{R}^d))$, we can choose $|\bar{T}|$ large enough such that
\begin{align*}
\|u_1-u_2\|_{V_{\bar{T}}}+\|v_1-v_2\|_{V_{\bar{T}}}\leq \frac{1}{2}[\|u_1-u_2\|_{V_{\bar{T}}}+\|v_1-v_2\|_{V_{\bar{T}}}]
\end{align*}
with
$$
V_{\bar{T}}=L^{\rho}(I,\dot{H}^{s,r}(\mathbb{R}^d))\cap L^{\infty}(I, L^{\infty}(I,\dot{H}^{1,2}(\mathbb{R}^d)),
$$
hence $u_1=u_2$, $v_1=v_2$ in $V_{\bar{T}}$. Step by step, it is possible to replace $\bar{T}$ by $\bar{T}+\epsilon$ with $\epsilon$ possibly depending on $u_1, u_2, v_1, v_2$, just like the arguments above, we can obtain $u_1=u_2$, $v_1=v_2$ in $V$. \hfill $\Box$

As a direct result of this theorem, we have the following corollary

{\bf Corollary 5.2} {\it
\begin{align}
\|u(t)-u^-_0(t)\|_e+\|v(t)-v^-_0(t)\|_e\rightarrow 0\quad {\rm as} \quad t\rightarrow -\infty.
\end{align}
}
Here $\|w(t)\|_e$ denote the energy norm
$$
\|w(\cdot,t)\|_e^2=\frac{1}{2}[\|(-\Delta)^{\frac{1}{2}}w(\cdot,t)\|^2_{L^2}+\|w(\cdot,t)\|^2_{L^2}].
$$

{\bf Proof:} Note that $L^{\rho}(\mathbb{R}, \dot{H}^{s,r}(\mathbb{R}^d))\subset L^{\rho}(\mathbb{R},L^{2\rho}(\mathbb{R}^d))$, we have
\begin{align*}
\|u(t)-u^-_0(t)\|_e+\|v(t)-v^-_0(t)\|_e&\leq c\int^t_{-\infty} [\||u|^{\alpha}|v|^{\beta+2}u(\tau)\|_{L^2}+\||u|^{\alpha+2}|v|^{\beta}v(\tau)\|_{L^2}]d\tau\nonumber\\
&\leq c\int^t_{-\infty}[\|u(\tau)\|^{\rho}_{L^{2\rho}}+\|v(\tau)\|^{\rho}_{L^{2\rho}}]d\tau\rightarrow 0
\end{align*}
as $t\rightarrow -\infty$. \hfill$\Box$

Define
\begin{align*}
u^+_0(t):=u(t)+\int_t^{\infty}(-\Delta)^{-\frac{1}{2}}\sin[(-\Delta)^{\frac{1}{2}}(t-\tau)](\lambda |u|^{\alpha}|v|^{\beta+2}u)d\tau,\quad u^+_0(0)=u^+,\ u^+_{0t}(0)=\tilde{u}^+\\
v^+_0(t):=v(t)+\int_t^{\infty}(-\Delta)^{-\frac{1}{2}}\sin[(-\Delta)^{\frac{1}{2}}(t-\tau)](\mu |u|^{\alpha+2}|v|^{\beta}v)d\tau,\quad v^+_0(0)=v^+,\ v^+_{0t}(0)=\tilde{v}^+.
\end{align*}
Exactly similar to the discussions above, we can get $\|u(t)-u^+_0(t)\|_e+\|v(t)-v^+_0(t)\|_e\rightarrow 0$ as $t\rightarrow +\infty$. Now we have proven

{\bf Theorem 5.3} {\it The scattering operators $S_1:(u_{10},u_{20})\rightarrow (u^+,\tilde{u}^+)$ and $S_2:(v_{10},v_{20})\rightarrow (v^+,\tilde{v}^+)$
exist in the sense of energy norms in a whole neighbourhood of the origin in $\dot{H}^{1,2}(\mathbb{R}^d)\times L^2(\mathbb{R}^d)$.
}

\section{$\dot{H}^{s_c}\times \dot{H}^{s_c-1}$ scattering results in the energy supercritical case}

\qquad In this section, we discuss the local wellposedness and scattering results on (\ref{1}) in the energy supercritical case. Denote $s_c=\frac{d}{2}-\frac{2}{\alpha+\beta+2}$. The space-time norm $\|\cdot\|_{L^1_tL^2_x}$ means that $\|\cdot\|_{L^1_tL^2_x((-T,T)\times \mathbb{R}^d)}$, while we will omit $(-T,T)\times \mathbb{R}^d)$ below, $0<T\leq +\infty$.
 We state the results as follows.

{\bf Theorem 6.1} {\it Let $d=3$ or $d=4$, $(u_{10},u_{20})\in \dot{H}^{s_c}(\mathbb{R}^d)\times \dot{H}^{s_c-1}(\mathbb{R}^d)$ and $(v_{10},v_{20})\in \dot{H}^{s_c}(\mathbb{R}^d)\times \dot{H}^{s_c-1}(\mathbb{R}^d)$.
Then the Cauchy problem (\ref{1}) possesses a unique solution $(u,v)$ with the maximal lifespan $u:I\times\mathbb{R}^d\rightarrow \mathbb{R}$ and
$v:I\times\mathbb{R}^d\rightarrow \mathbb{R}$.

Moreover, there exists $\delta_0>0$ and if
$$
\|u_{10}\|_{\dot{H}^{s_c}(\mathbb{R}^d)}+\|u_{20}\|_{\dot{H}^{s_c-1}(\mathbb{R}^d)}+\|v_{10}\|_{\dot{H}^{s_c}(\mathbb{R}^d)}+\|v_{20}\|_{\dot{H}^{s_c-1}(\mathbb{R}^d)}\leq \delta_0,
$$
there exist functions pairs  $(u_{1\pm},u_{2\pm})\in \dot{H}^{s_c}(\mathbb{R}^d)\times \dot{H}^{s_c-1}(\mathbb{R}^d)$ and $(v_{1\pm},v_{2\pm})\in \dot{H}^{s_c}(\mathbb{R}^d)\times \dot{H}^{s_c-1}(\mathbb{R}^d)$ such that
\begin{equation}
\label{10031} \left\|\left[
  \begin{array}{llll}
 u(t)\\
 u_t(t)
   \end{array}
    \right]
  -
  \left[
 \begin{array}{llll}
 \cos((t-t_0)|\nabla|& |\nabla|^{-1}\sin(t-t_0)|\nabla|\\
-|\nabla|\sin(t-t_0)|\nabla|& \cos(t-t_0)|\nabla|
  \end{array}
    \right]\left[
  \begin{array}{llll}
 u_{1\pm}\\
 u_{2\pm}
   \end{array}
    \right]
 \right\|_{\dot{H}^{s_c}\times \dot{H}^{s_c-1}}\rightarrow 0.
\end{equation}
and
\begin{equation}
\label{10032} \left\|\left[
  \begin{array}{llll}
 v(t)\\
 v_t(t)
   \end{array}
    \right]
  -
  \left[
 \begin{array}{llll}
 \cos((t-t_0)|\nabla|& |\nabla|^{-1}\sin(t-t_0)|\nabla|\\
-|\nabla|\sin(t-t_0)|\nabla|& \cos(t-t_0)|\nabla|
  \end{array}
    \right]\left[
  \begin{array}{llll}
 v_{1\pm}\\
 v_{2\pm}
   \end{array}
    \right]
 \right\|_{\dot{H}^{s_c}\times \dot{H}^{s_c-1}}\rightarrow 0.
\end{equation}
as $t\rightarrow \pm \infty$. }

{\bf Proof:}  First, applying Strichartz's estimate to (\ref{771}), we have
\begin{align}
\|u\|_{L^{\infty}_t\dot{H}^{s_c}_x}+\|v\|_{L^{\infty}_t\dot{H}^{s_c}_x}&\lesssim \|u_{10}\|_{\dot{H}^{s_c}_x}+\|u_{20}\|_{\dot{H}^{s_c-1}_x}+\||\nabla|^{s_c-1}(|u|^{\alpha}|v|^{\beta+2}u)\|_{L^1_tL^2_x}\nonumber\\
&\quad+\|v_{10}\|_{\dot{H}^{s_c}_x}+\|v_{20}\|_{\dot{H}^{s_c-1}_x}+\||\nabla|^{s_c-1}(|u|^{\alpha+2}|v|^{\beta}v)\|_{L^1_tL^2_x}.\label{8302}
\end{align}

Note that $s_c<2$. We discuss it in two cases.

{\bf Case 1: $d=3$.}  We define the working space as follows.
\begin{align}
\|u\|_{X_T}+\|v\|_{X_T}&=\|u\|_{L^{\infty}_t\dot{H}^{s_c}_x}+\||\nabla|^{s_c-1}u\|_{L^{2^+}L^{\infty-}_x}+\|u\|_{L^{2(\alpha+\beta+2)-}_tL^{2(\alpha+\beta+2)-}_x}\nonumber\\
&\quad+\|v\|_{L^{\infty}_t\dot{H}^{s_c}_x}+\||\nabla|^{s_c-1}v\|_{L^{2^+}L^{\infty-}_x}+\|v\|_{L^{2(\alpha+\beta+2)-}_tL^{2(\alpha+\beta+2)-}_x}.\label{8301}
\end{align}

By Lemma 2.5 and Lemma 2.6, we obtain
\begin{align}
&\quad \||\nabla|^{s_c-1}(|u|^{\alpha}|v|^{\beta+2}u)\|_{L^1_tL^2_x}+\||\nabla|^{s_c-1}(|u|^{\alpha+2}|v|^{\beta}v)\|_{L^1_tL^2_x}\nonumber\\
&\lesssim \left\|[\|u\|^{\alpha+\beta+2}_{L^{2(\alpha+\beta+2)+}_x}+\|v\|^{\alpha+\beta+2}_{L^{2(\alpha+\beta+2)+}_x} ]  [\||\nabla|^{s_c-1}u\|_{L^{\infty-}_x}+\||\nabla|^{s_c-1}v\|_{L^{\infty-}_x}]\right\|_{L^1_t}\nonumber\\
&\lesssim [\|u\|^{\alpha+\beta+2}_{L^{2(\alpha+\beta+2)-}_tL^{2(\alpha+\beta+2)+}_x}+\|v\|^{\alpha+\beta+2}_{L^{2(\alpha+\beta+2)-}_tL^{2(\alpha+\beta+2)+}_x} ]  [\||\nabla|^{s_c-1}u\|_{L^{2+}_tL^{\infty-}_x}+\||\nabla|^{s_c-1}v\|_{L^{2+}_tL^{\infty-}_x}]\nonumber\\
&\lesssim \|u\|^{\alpha+\beta+3}_{X_T}+\|v\|^{\alpha+\beta+3}_{X_T}.\label{8311}
\end{align}

Letting $(a',b')=(1,2)$ in Lemma 2.4, using Strichartz's estimate, we get
\begin{align}
&\quad\||\nabla|^{s_c-1} u\|_{L^{2+}_tL^{\infty-}_x}+\||\nabla|^{s_c-1} v\|_{L^{2+}_tL^{\infty-}_x}\nonumber\\
&\lesssim \|u_{10}\|_{\dot{H}^{s_c}_x}+\|u_{20}\|_{\dot{H}^{s_c-1}_x}+\||\nabla|^{s_c-1}(|u|^{\alpha}|v|^{\beta+2}u)\|_{L^1_tL^2_x}\nonumber\\
&\quad+\|u_{10}\|_{\dot{H}^{s_c}_x}+\|u_{20}\|_{\dot{H}^{s_c-1}_x}+\||\nabla|^{s_c-1}(|u|^{\alpha+2}|v|^{\beta}v)\|_{L^1_tL^2_x}\label{8312}
\end{align}
and
\begin{align}
&\quad\| u\|_{L^{2(\alpha+\beta+2)-}_tL^{2(\alpha+\beta+2)+}_x}+\||v\|_{L^{2(\alpha+\beta+2)-}_tL^{2(\alpha+\beta+2)+}_x}\nonumber\\
&\lesssim \|u_{10}\|_{\dot{H}^{s_c}_x}+\|u_{20}\|_{\dot{H}^{s_c-1}_x}+\||\nabla|^{s_c-1}(|u|^{\alpha}|v|^{\beta+2}u)\|_{L^1_tL^2_x}\nonumber\\
&\quad+\|u_{10}\|_{\dot{H}^{s_c}_x}+\|u_{20}\|_{\dot{H}^{s_c-1}_x}+\||\nabla|^{s_c-1}(|u|^{\alpha+2}|v|^{\beta}v)\|_{L^1_tL^2_x}.\label{8313}
\end{align}
Putting the results of (\ref{8301})--(\ref{8313}) together, we have
\begin{align*}
\|u\|_{X_T}+\|v\|_{X_T}\lesssim \|u_{10}\|_{\dot{H}^{s_c}_x}+\|u_{20}\|_{\dot{H}^{s_c-1}_x}+\|v_{10}\|_{\dot{H}^{s_c}_x}+\|v_{20}\|_{\dot{H}^{s_c-1}_x}+\|u\|^{\alpha+\beta+3}_{X_T}+\|v\|^{\alpha+\beta+3}_{X_T}.
\end{align*}
If $T$ small enough, we have
\begin{align}
\|u\|_{X_T}+\|v\|_{X_T}\leq C[\|u_{10}\|_{\dot{H}^{s_c}_x}+\|u_{20}\|_{\dot{H}^{s_c-1}_x}+\|v_{10}\|_{\dot{H}^{s_c}_x}+\|v_{20}\|_{\dot{H}^{s_c-1}_x}]:=M \label{911}
\end{align}
for some $M$ large enough.

{\bf Case 2: $d=4$.} We define the working space as
\begin{align}
\|u\|_{X_T}+\|v\|_{X_T}&=\|u\|_{L^{\infty}_t\dot{H}^{s_c}_x}+\||\nabla|^{s_c-1}u\|_{L^2_tL^8_x}+\|u\|_{L^{2(\alpha+\beta+2)}_tL^{\frac{8(\alpha+\beta+2)}{3}}_x}\nonumber\\
&\quad+\|u\|_{L^{\infty}_t\dot{H}^{s_c}_x}+\||\nabla|^{s_c-1}u\|_{L^2_tL^8_x}+\|u\|_{L^{2(\alpha+\beta+2)}_tL^{\frac{8(\alpha+\beta+2)}{3}}_x}.\label{91x1}
\end{align}

By Lemma 2.5 and Lemma 2.6, we obtain
\begin{align}
&\quad \||\nabla|^{s_c-1}(|u|^{\alpha}|v|^{\beta+2}u)\|_{L^1_tL^2_x}+\||\nabla|^{s_c-1}(|u|^{\alpha+2}|v|^{\beta}v)\|_{L^1_tL^2_x}\nonumber\\
&\lesssim \left\|[\|u\|^{\alpha+\beta+2}_{L^{\frac{8(\alpha+\beta+2)}{3}}_x}+\|v\|^{\alpha+\beta+2}_{L^{\frac{8(\alpha+\beta+2)}{3}}_x} ]  [\||\nabla|^{s_c-1}u\|_{L^8_x}+\||\nabla|^{s_c-1}v\|_{L^8_x}]\right\|_{L^1_t}\nonumber\\
&\lesssim [\|u\|^{\alpha+\beta+2}_{L^{2(\alpha+\beta+2)}_tL^{\frac{8(\alpha+\beta+2)}{3}}_x}+\|v\|^{\alpha+\beta+2}_{L^{2(\alpha+\beta+2)}_tL^{\frac{8(\alpha+\beta+2)}{3}}_x} ]  [\||\nabla|^{s_c-1}u\|_{L^2_tL^8_x}+\||\nabla|^{s_c-1}v\|_{L^2_tL^8_x}]\nonumber\\
&\lesssim \|u\|^{\alpha+\beta+3}_{X_T}+\|v\|^{\alpha+\beta+3}_{X_T}.\label{921}
\end{align}

Letting $(a',b')=(1,2)$ in Lemma 2.4, using Strichartz's estimate, we get
\begin{align}
&\quad\||\nabla|^{s_c-1} u\|_{L^2_tL^8_x}+\||\nabla|^{s_c-1} v\|_{L^2_tL^8_x}\nonumber\\
&\lesssim \|u_{10}\|_{\dot{H}^{s_c}_x}+\|u_{20}\|_{\dot{H}^{s_c-1}_x}+\||\nabla|^{s_c-1}(|u|^{\alpha}|v|^{\beta+2}u)\|_{L^1_tL^2_x}\nonumber\\
&\quad+\|u_{10}\|_{\dot{H}^{s_c}_x}+\|u_{20}\|_{\dot{H}^{s_c-1}_x}+\||\nabla|^{s_c-1}(|u|^{\alpha+2}|v|^{\beta}v)\|_{L^1_tL^2_x}.\label{922}
\end{align}
and
\begin{align}
&\quad\| u\|_{L^{2(\alpha+\beta+2)}_tL^{\frac{8(\alpha+\beta+2)}{3}}_x}+\||v\|_{L^{2(\alpha+\beta+2)}_tL^{\frac{8(\alpha+\beta+2)}{3}}_x}\nonumber\\
&\lesssim \|u_{10}\|_{\dot{H}^{s_c}_x}+\|u_{20}\|_{\dot{H}^{s_c-1}_x}+\||\nabla|^{s_c-1}(|u|^{\alpha}|v|^{\beta+2}u)\|_{L^1_tL^2_x}\nonumber\\
&\quad+\|u_{10}\|_{\dot{H}^{s_c}_x}+\|u_{20}\|_{\dot{H}^{s_c-1}_x}+\||\nabla|^{s_c-1}(|u|^{\alpha+2}|v|^{\beta}v)\|_{L^1_tL^2_x}.\label{923}
\end{align}
Putting the results of (\ref{91x1})--(\ref{923}) together, we obtain
\begin{align*}
\|u\|_{X_T}+\|v\|_{X_T}\lesssim \|u_{10}\|_{\dot{H}^{s_c}_x}+\|u_{20}\|_{\dot{H}^{s_c-1}_x}+\|v_{10}\|_{\dot{H}^{s_c}_x}+\|v_{20}\|_{\dot{H}^{s_c-1}_x}+\|u\|^{\alpha+\beta+3}_{X_T}+\|v\|^{\alpha+\beta+3}_{X_T}.
\end{align*}
If $T$ small enough, we have
\begin{align}
\|u\|_{X_T}+\|v\|_{X_T}\leq C[\|u_{10}\|_{\dot{H}^{s_c}_x}+\|u_{20}\|_{\dot{H}^{s_c-1}_x}+\|v_{10}\|_{\dot{H}^{s_c}_x}+\|v_{20}\|_{\dot{H}^{s_c-1}_x}]:=M\label{924'}
\end{align}
for some $M$ large enough.

Let ${\bf \mathcal{A}}(u,v)=(A_1(u_0,v_0), A_2(u_0,v_0))$ and $B_0(M)$ is the ball in $X_T\times X_T$ with the center is $(0,0)$ and the radium is $M$. Then (\ref{924'}) means that ${\bf \mathcal{A}}$  reflects  the ball
$B_0(M)$ to itself. Moreover, ${\bf \mathcal{A}}$  is contract and Banach's fixed point theorem gives a unique solution of (\ref{1}).

Using the bootstrap argument, if $\|u_{10}\|_{\dot{H}^{s_c}_x}+\|u_{20}\|_{\dot{H}^{s_c-1}_x}+\|v_{10}\|_{\dot{H}^{s_c}_x}+\|v_{20}\|_{\dot{H}^{s_c-1}_x}<\delta_0$ small enough, we have
\begin{align}
\|u\|_{X_T}+\|v\|_{X_T}\lesssim \delta_0,\label{924}
\end{align}
 which implies that
\begin{align*}
\|u\|_{X_{\infty}}+\|v\|_{X_{\infty}}\lesssim \delta_0.
\end{align*}

Now we choose scattering states as
\begin{equation}
\left[
  \begin{array}{llll}
 u_{1\pm}\\
 u_{2\pm}
   \end{array}
    \right]
  =
  \left[\begin{array}{llll}
 u_{10}\\
 u_{20}
   \end{array}
    \right]
    -\int^{+\infty}_0
  \left[
 \begin{array}{llll}
 |\nabla|^{-1}\sin(-s|\nabla|)\\
\cos(-s|\nabla|)
  \end{array}
    \right](|u(s)|^{\alpha}|v(s)|^{\beta+2}u(s)ds\label{9191}
\end{equation}
and
\begin{equation}
\left[
  \begin{array}{llll}
 v_{1\pm}\\
 v_{2\pm}
   \end{array}
    \right]
  =
  \left[\begin{array}{llll}
 v_{10}\\
 v_{20}
   \end{array}
    \right]
    -\int^{+\infty}_0
  \left[
 \begin{array}{llll}
 |\nabla|^{-1}\sin(-s|\nabla|)\\
\cos(-s|\nabla|)
  \end{array}
    \right](|u(s)|^{\alpha+2}|v(s)|^{\beta}v(s)ds\label{9192}
\end{equation}
Therefore,
\begin{align}
&\quad \left\|\left[
  \begin{array}{llll}
 u(t)\\
 u_t(t)
   \end{array}
    \right]
  -
  \left[
 \begin{array}{llll}
 \cos(t-t_0)|\nabla|& |\nabla|^{-1}\sin(t-t_0)|\nabla|\\
-|\nabla|\sin(t-t_0)|\nabla|& \cos(t-t_0)|\nabla|
  \end{array}
    \right]\left[
  \begin{array}{llll}
 u_{1\pm}\\
 u_{2\pm}
   \end{array}
    \right]
 \right\|_{\dot{H}^{s_c}\times \dot{H}^{s_c-1}}\nonumber\\
&=\|{\bf A}(t)\int_t^{\infty}
  \left[
 \begin{array}{llll}
 |\nabla|^{-1}\sin(-s|\nabla|)\\
\cos(-s|\nabla|)
  \end{array}
    \right](|u(s)|^{\alpha+2}|v(s)|^{\beta}v(s))ds\|_{\dot{H}^{s_c}\times \dot{H}^{s_c-1}}\label{9193}
\end{align}
Here
\begin{align}
\bf{A}(t)=\left[
 \begin{array}{llll}
 \cos(t-t_0)|\nabla|& |\nabla|^{-1}\sin(t-t_0)|\nabla|\\
-|\nabla|\sin(t-t_0)|\nabla|& \cos(t-t_0)|\nabla|
  \end{array}
    \right].\label{9194}
\end{align}
Using Strichartz estimate, we can obtain
\begin{align}
\label{10033}&\quad \left\|\left[
  \begin{array}{llll}
 u(t)\\
 u_t(t)
   \end{array}
    \right]
  -
  \left[
 \begin{array}{llll}
 \cos(t-t_0)|\nabla|& |\nabla|^{-1}\sin(t-t_0)|\nabla|\\
-|\nabla|\sin(t-t_0)|\nabla|& \cos(t-t_0)|\nabla|
  \end{array}
    \right]\left[
  \begin{array}{llll}
 u_{1\pm}\\
 u_{2\pm}
   \end{array}
    \right]
 \right\|_{\dot{H}^{s_c}\times \dot{H}^{s_c-1}}\nonumber\\
&\lesssim \||\nabla|^{s_c-1}(|u(s)|^{\alpha+2}|v(s)|^{\beta}v(s))\|_{L^{a'}_tL^{b'}_x((t,+\infty)\times \mathbb{R}^N)}\nonumber\\
&\lesssim \|u\|^{\alpha+\beta+3}_{X_{(t,+\infty)}}+\|v\|^{\alpha+\beta+3}_{X_{(t,+\infty)}}\rightarrow 0\quad {\rm as} \quad t\rightarrow +\infty,
\end{align}
which means (\ref{10031}). Here
\begin{align}
\frac{1}{a}+\frac{1}{a'}=1,\quad \frac{1}{b}+\frac{1}{b'}=1\label{10034}
\end{align}
and
\begin{align}
\frac{1}{a'}+\frac{N}{b'}=\frac{N+2}{2}.\label{10035}
\end{align}

(\ref{10032}) can be obtained similarly.\hfill $\Box$

{\bf  Declarations}

All authors declare no conflicts of interest in this paper.

\end{document}